\RequirePackage{fix-cm}
\documentclass[smallextended]{svjour3}       
\smartqed  
\usepackage[latin1]{inputenc}
\usepackage{amsmath}
\usepackage[ngerman,english]{babel}
\usepackage{amsfonts}
\usepackage{amssymb}
\usepackage{mathtools}
\usepackage[bookmarks]{hyperref}
\usepackage{graphicx,xcolor}
\usepackage{dsfont}
\usepackage{caption}
\usepackage{subcaption}
\usepackage{array,multirow}
\usepackage{tikz}
\usetikzlibrary{decorations.pathreplacing}

\newcommand{\expval}[1]{\left\langle #1\right\rangle}

\newcommand{\leff}{\lambda_\text{eff}}
\newcommand{\wt}[1]{\widetilde{#1}}

\newcommand{\stot}{\expval{\Delta s_\text{tot}}}

\newcommand{\tcross}{t^{\text{EW}\to\text{KPZ}}}
\newcommand{\tcorr}{t_\text{c}^\text{KPZ}}
\newcommand{\tcEW}{t_\text{c}^\text{EW}}
\newcommand{\tp}{t^\prime}
\newcommand{\tpp}{t^{\prime\prime}}
\newcommand{\qp}{q^\prime}
\newcommand{\op}{\omega^\prime}
\newcommand{\Lp}{\Lambda^\prime}
\newcommand{\Lo}{\Lambda_0}

\begin{document}

\title{The Two Scaling Regimes of the Thermodynamic Uncertainty Relation for the KPZ--Equation}

\titlerunning{The Two Scaling Regimes of the KPZ--TUR}

\author{Oliver Niggemann  \and  Udo Seifert}

\institute{O. Niggemann \at
              II. Institute for Theoretical Physics, University of Stuttgart, Pfaffenwaldring 57, 70550 Stuttgart, Germany \\
              \email{niggemann@theo2.physik.uni-stuttgart.de}
           \and
           U. Seifert \at
              II. Institute for Theoretical Physics, University of Stuttgart, Pfaffenwaldring 57, 70550 Stuttgart, Germany\\
              \email{useifert@theo2.physik.uni-stuttgart.de}
}

\date{Received: date / Accepted: date}

\maketitle

\begin{abstract}
We investigate the thermodynamic uncertainty relation for the $(1+1)$ dimensional Kardar-Parisi-Zhang equation on a finite spatial interval. In particular, we extend the results for small coupling strengths obtained previously to large values of the coupling parameter. It will be shown that, due to the scaling behavior of the KPZ equation, the TUR product displays two distinct regimes which are separated by a critical value of an effective coupling parameter. The asymptotic behavior below and above the critical threshold is explored analytically. For small coupling, we determine this product perturbatively including the fourth order; for strong coupling we employ a dynamical renormalization group approach. Whereas the TUR product approaches a value of $5$ in the weak coupling limit, it asymptotically displays a linear increase with the coupling parameter for strong couplings. The analytical results are then compared to direct numerical simulations of the KPZ equation showing convincing agreement.
\end{abstract}

\pagebreak

\section{Introduction}\label{sec:Intro}

Over the last years there has been remarkable progress in field theory with regard to the Kardar-Parisi-Zhang (KPZ) dynamics \cite{KPZ1986} on the one hand and in stochastic thermodynamics with respect to the thermodynamic uncertainty relation (TUR) on a discrete set of states \cite{BaratoSeifert2015,Gingrich2016,GingrichReview2020} on the other hand. The KPZ equation is a paradigmatic example of a growth equation displaying non-equilibrium dynamics while the TUR bounds the entropy production through fluctuation and mean of any current. For a recent excerpt of the former see, e.g., \cite{HalpinHealyTakeuchi2015,Takeuchi2017,Spohn2020,Sasamoto2016,Praehofer2004,Amir2011,Takeuchi2012,Imamura2012,Krug2019,Meerson2018,Canet2012,Canet2021,Fukai2017,Fukai2020,Iwatsuka2020}. In regard of the latest achievements for the TUR see, e.g., \cite{Pal2021,Terlizzi2020,Koyuk2020,Liu2020,Koyuk2019}. In \cite{NiggemannSeifert2020,NiggemannSeifert2021} these two areas have been connected by formulating a general field-theoretic TUR and evaluating it explicitly for the KPZ equation analytically as well as numerically in a certain scaling regime. For other field-theoretic formulations of stochastic thermodynamic concepts, in particular the stochastic entropy production, see, e.g., \cite{Wio2019,Cates2017,Cates2021}.\\
The derivation of the KPZ-TUR in \cite{NiggemannSeifert2020} relies on a perturbational approximation in a small effective coupling parameter of the KPZ non-linearity. This approach is quite generally applicable to stochastic field-theoretic overdamped Langevin equations. However, it is intrinsically limited to the linear scaling regime of the respective partial differential equation \cite{HentschelFamily1991}. In case of the KPZ equation, this is the so-called Edwards-Wilkinson (EW) scaling regime \cite{EW1982,KrugReview1997}. The aim of the present paper is to extend the results from \cite{NiggemannSeifert2020} valid in the EW scaling regime of the KPZ equation to the genuine KPZ scaling regime. This requires an approach that will hold for arbitrary values of the effective coupling strength of the KPZ non-linearity. For equal-time correlation functions such a generalization is possible by using the exactly known stationary probability density functional of the $1d$ KPZ equation. For two-time correlation functions, however, this approach is not feasible and thus different methods have to be used. In the present case we use two different ways of calculating this type of correlation functions. The first one is the perturbational approximation introduced in \cite{NiggemannSeifert2020} and as a second one we employ the dynamic renormalization group (DRG) approach. Here the former applies to the EW scaling regime, where the latter covers the genuine KPZ scaling regime. Hence, a combination of these methods enables us to analytically express the KPZ-TUR for arbitrary values of the effective coupling parameter. These results are compared with numerical simulations based on the method from \cite{NiggemannSeifert2021}. This comparison will show convincing agreement between the theoretical predictions and the numerical results.\\
The paper is organized as follows. In \autoref{sec:ProblemState} we give a brief overview of the problem at hand and introduce the necessary notions for the formulation of the KPZ-TUR. Section \ref{sec:ExactResults} deals with the derivation of exact results valid for arbitrary coupling strength. In particular, we utilize the stationary state probability density functional of the $(1+1)$ dimensional KPZ equation to calculate equal-time correlation functions entering the KPZ-TUR via functional integration. In \autoref{sec:ScalingVar} we concisely state the scaling behavior of the KPZ equation as this will be relevant for the calculation of temporal two-point correlation functions. Sections \ref{sec:PerturbationExpansion} and \ref{sec:RGCalculation} cover the calculation of a specific two-time correlation function via perturbational approximation and DRG, respectively. The combination of the results obtained in the prior sections, yields the KPZ-TUR for arbitrary coupling strength, which is given in \autoref{sec:TURFull}. The comparison of the analytically obtained theoretical predictions to numerical data is shown in \autoref{sec:CompNumSim}. We summarize our results in \autoref{sec:Conclusion}.

\section{The Problem}\label{sec:ProblemState}

In this section we will briefly introduce the KPZ equation and the TUR, as well as give a short summary of the results obtained in \cite{NiggemannSeifert2020} which link the two topics. At the end of the section we outline the steps to be taken in order to extend the results from \cite{NiggemannSeifert2020,NiggemannSeifert2021} to arbitrary coupling strength.\\
We begin with stating the KPZ equation in the form needed for our analysis, i.e., the $(1+1)$ dimensional Kardar-Parisi-Zhang equation on a finite interval $x\in[0,b]$, $b>0$, given by
\begin{equation}
\partial_th(x,t)=\nu\partial_x^2h(x,t)+\frac{\lambda}{2}\left(\partial_xh(x,t)\right)^2+\eta(x,t).\label{eq:KPZEq}
\end{equation}
Here $\nu$ represents the surface tension, $\lambda$ is the coupling parameter of the non-linearity and $\eta$ Gaussian space-time white noise with zero mean and autocorrelation $\expval{\eta(x,t)\eta(x^\prime,t^\prime)}=\Delta_0\delta(x-x^\prime)\delta(t-t^\prime)$, $\Delta_0$ denoting the noise strength. We further assume periodic boundary conditions for \eqref{eq:KPZEq}, i.e., $h(0,t)=h(b,t)$, and flat initial condition $h(x,0)=0$ (see also \cite{NiggemannSeifert2020}).\\
 The thermodynamic uncertainty relation for a non-equilibrium steady state (NESS) was originally proposed for Markovian networks \cite{BaratoSeifert2015}. It gives a lower bound on the total entropy production $\stot$ needed to provide a certain precision $\epsilon^2$ of a process in such a network. It reads
\begin{equation}
\mathcal{Q}\equiv\stot\,\epsilon^2\geq2,\label{eq:TURGen}
\end{equation}
where $\expval{\cdot}$ denotes averages with respect to the noise history. Here, $\stot=\sigma\,t$ in the stationary state, with $\sigma$ the entropy production rate and $\epsilon^2=2D/(j^2t)$, with $D$ the diffusivity and $j$ an arbitrary NESS current. Later, the TUR \eqref{eq:TURGen} was proven to hold for a Markovian dynamics on a discrete set of states \cite{Gingrich2016} and for overdamped Langevin dynamics \cite{GingrichRotskoff2017}.\\
Recently, the TUR in \eqref{eq:TURGen} was extended to a general field-theoretic overdamped Langevin equation \cite{NiggemannSeifert2020} and exemplified with the $(1+1)$ dimensional KPZ equation from \eqref{eq:KPZEq}. For the KPZ equation it was found via a perturbative calculation of $\stot$ and $\epsilon^2$ as well as by direct numerical simulation \cite{NiggemannSeifert2021} that $\mathcal{Q}\simeq5$ for small values of the effective coupling parameter
\begin{equation}
\leff\equiv\sqrt{\frac{\Delta_0\,b}{\nu^3}}\,\lambda.\label{eq:DefLambdaEff}
\end{equation}
What is meant by \lq small\rq{} will be specified below in \autoref{sec:ScalingVar}. According to \cite{HentschelFamily1991}, such a perturbative approach to a non-linear PDE like \eqref{eq:KPZEq} will yield results expected to be valid in the linear scaling regime of the non-linear equation. Hence, the results from \cite{NiggemannSeifert2020,NiggemannSeifert2021} are valid in the so-called Edwards-Wilkinson (EW) scaling regime of the KPZ equation.\\
In the present paper, we will derive the field-theoretic analog of \eqref{eq:TURGen} (see \cite{NiggemannSeifert2020}) for arbitrary values of $\leff$ and thus extend the range of validity from the EW scaling regime to the genuine KPZ scaling regime. The terminology will be explained in more detail in \autoref{sec:ScalingVar}.\\
The expressions for the constituents of the TUR used in this paper are derived in \cite{NiggemannSeifert2020} and read
\begin{equation}
\stot\equiv\frac{\lambda^2}{2\,\Delta_0}\int_0^td\tau\,\expval{\int_0^bdx\,\left(\partial_xh(x,\tau)\right)^4}\label{eq:DefStot}
\end{equation}
as the total entropy production and
\begin{equation}
\epsilon^2\equiv\frac{\text{var}\left[\Psi_g(t)\right]}{\expval{\Psi_g(t)}^2}=\frac{\expval{\left(\Psi_g(t)-\expval{\Psi_g(t)}\right)^2}}{\expval{\Psi_g(t)}^2},\label{eq:DefPrecision_g}
\end{equation}
as the precision, where
\begin{equation}
\Psi_g(t)\equiv\int_0^bdx\,g(x)\,h(x,t).\label{eq:DefPsi_g}
\end{equation}
Here, $\Psi_g$ describes the time-integrated generalized current with $g(x)\in L^2(0,b)$ ($\int_0^bdx\,g(x)\neq0$) as an arbitrary weight function. As it was shown in \cite{NiggemannSeifert2020} that $\mathcal{Q}$ does not depend on the choice of $g(x)$, we will set $g=1$ in the following, i.e.,
\begin{equation}
\Psi(t)\equiv\Psi_1(t)=\int_0^bdx\,h(x,t).\label{eq:DefPsi}
\end{equation}
Hence, \eqref{eq:DefPrecision_g} becomes
\begin{equation}
\epsilon^2=\frac{\text{var}\left[\Psi(t)\right]}{\expval{\Psi(t)}^2}=\frac{\expval{\left(\Psi(t)-\expval{\Psi(t)}\right)^2}}{\expval{\Psi(t)}^2},\label{eq:DefPrecision}
\end{equation}
with $\Psi(t)$ from \eqref{eq:DefPsi}. In the stationary state we have $\expval{\Psi(t)}=J\,t$ with $J$ the stationary current.\\
In the following we derive explicit expressions for $\stot$, $\expval{\Psi(t)}$ and $\text{var}[\Psi(t)]$. The first two, namely $\stot$ and $\expval{\Psi(t)}$, are given by equal-time correlation functions. These correlation functions may thus be calculated in the stationary state via functional integration over the stationary state probability density of the $(1+1)$ dimensional KPZ equation (see \autoref{sec:ExactResults}). The variance of $\Psi(t)$ is, on the other hand, given by a temporal two-point correlation function, which requires more knowledge than the stationary state probability density. We show below two different ways to obtain $\text{var}[\Psi(t)]$. The first uses a perturbation expansion in $\leff$ from \eqref{eq:DefLambdaEff} (see \autoref{sec:PerturbationExpansion}) and the second employs dynamic renormalization group (DRG) techniques (see \autoref{sec:RGCalculation}). 

\section{Exact Results}\label{sec:ExactResults}

\subsection{Normalized Stationary Distribution}\label{subsec:ER_NormalizedPsH}

For the $(1+1)$ dimensional KPZ equation the stationary probability density functional of the height field $h(x,t)$ is known exactly \cite{Takeuchi2017,KrugReview1997,HalpinHealyZhang1995} and reads
\begin{equation}
p_\text{s}[h]\sim\exp\left[-\frac{\nu}{\Delta_0}\int_0^bdx\,\left(\partial_xh\right)^2\right].\label{eq:DefPsOfH}
\end{equation}
Note, that \eqref{eq:DefPsOfH} is identical to the steady state solution of the Fokker-Planck equation for the linear problem, i.e., for the EW equation \cite{EW1982}. In the following, we want to use \eqref{eq:DefPsOfH} to calculate equal-time steady-state correlation functions. Hence, \eqref{eq:DefPsOfH} needs to be properly normalized. The normalization is obtained by expressing $h(x,t)$ in terms of its Fourier series (see e.g. \cite{NiggemannSeifert2020})
\begin{equation}
h(x,t)=\sum_{k\in\mathcal{R}}h_k(t)\,e^{2\pi i kx/b},\qquad\mathcal{R}=[-\Lambda,\Lambda],\,\Lambda\in\mathds{N},\label{eq:DefFourierSeriesH}
\end{equation}
where $h_k(t)\in\mathds{C}$, and inserting \eqref{eq:DefFourierSeriesH} into \eqref{eq:DefPsOfH}. The introduction of a finite Fourier-cutoff $\Lambda$ ensures the normalizability of \eqref{eq:DefPsOfH}. A subsequent functional integration of \eqref{eq:DefPsOfH} over $h$ yields
\begin{align}
\begin{split}
&\int\mathcal{D}[h]\,\exp\left[-\frac{\nu}{\Delta_0}\int_0^bdx\,\left(\partial_xh\right)^2\right]\\
&=\prod_{k=1}^\Lambda\int dh_{R,k}\,\exp\left[-\frac{8\pi^2\nu}{\Delta_0b}k^2h^2_{R,k}\right]\prod_{l=1}^\Lambda\int dh_{I,l}\,\exp\left[-\frac{8\pi^2\nu}{\Delta_0b}l^2h^2_{I,l}\right]\\
&=\left(\frac{\Delta_0\,b}{8\pi\,\nu}\right)^\Lambda\left(\frac{1}{\Lambda !}\right)^2,
\end{split}\label{eq:CalculationNormalizationPsH}
\end{align}
where $h_{R/I,j}(t)$ represents the real/imaginary part of $h_j(t)$, respectively. Hence, the normalization of \eqref{eq:DefPsOfH} reads
\begin{equation}
\mathcal{N}=\left(\frac{\Delta_0\,b}{8\pi\,\nu}\right)^\Lambda\left(\frac{1}{\Lambda !}\right)^2,\label{eq:DefNormalization}
\end{equation}
and therefore
\begin{equation}
p_\text{s}[h]=\frac{1}{\mathcal{N}}\,\exp\left[-\frac{\nu}{\Delta_0}\int_0^bdx\,\left(\partial_xh\right)^2\right].\label{eq:PsOfHNormalized}
\end{equation}
With \eqref{eq:PsOfHNormalized} we can explicitly calculate steady state equal-time correlation functions of the Fourier coefficients $h_k(t)$ by a functional integration, where $\expval{\cdot}_{p_\text{s}[h]}\equiv\int\mathcal{D}[h]\,(\cdot)p_\text{s}[h]$ is understood as the expectation value with respect to the stationary probability distribution. In particular
\begin{equation}
\expval{h_k(t)\,h_l(t)}_{p_\text{s}[h]}=-\frac{\Delta_0\,b}{\mu_k+\mu_l}\,\delta_{k,-l},\label{eq:DefCorrFuncHksPsH}
\end{equation}
with $\mu_k=-4\pi^2\nu k^2$ (see also \cite{NiggemannSeifert2020}). Note, that
\begin{equation}
\lim_{t\to\infty}\expval{(\cdot)(t)}=\expval{(\cdot)(t)}_{p_\text{s}[h]},\label{eq:ConnectionNoiseAveragePsH}
\end{equation}
is expected to hold, where $\expval{\cdot}$ denotes averages with respect to the noise history. We will show this explicitly in the case of $\expval{\Psi(t)}$ and $\stot$ below.

\subsection{Exact Stationary Current and Entropy Production Rate}\label{subsec:ER_Current_Sigma}

For the steady state current $J$, where $\expval{\Psi(t)}=J\,t$, we get
\begin{equation}
J=\expval{\partial_t\Psi(t)}_{p_\text{s}[h]}=\frac{\lambda}{2}\int_0^bdx\,\expval{\left(\partial_xh(x,t)\right)^2}_{p_\text{s}[h]}=\frac{\Delta_0\,\lambda}{2\,\nu}\,\Lambda.\label{eq:ExactSSCurrent}
\end{equation}
The second step follows from a spatial integration of \eqref{eq:KPZEq} with a subsequent averaging with respect to $p_\text{s}[h]$ and the last step uses Parseval's identity and \eqref{eq:DefCorrFuncHksPsH}. The result in \eqref{eq:ExactSSCurrent} has already been derived in \cite{NiggemannSeifert2020} as lowest order approximation of a perturbation expansion in $\leff$ where the l.h.s. of \eqref{eq:ConnectionNoiseAveragePsH} was used for calculating expectation values. It is instructive to examine why the lowest order approximation is in fact exact. This can best be seen by studying the structure of the perturbation expansion of $\expval{\partial_t\Psi(t)}$. Terms with an even power of $\leff$ vanish as they represent odd moments of the Gaussian noise $\eta$, whereas terms with odd power greater than $1$ vanish by exact cancellation of the involved moments.\\
For the steady-state entropy production rate $\sigma$, with $\stot=\sigma\,t$, it is found with \eqref{eq:DefFourierSeriesH}, using Wick's theorem and \eqref{eq:DefCorrFuncHksPsH} that
\begin{align}
\begin{split}
\sigma&=\frac{\lambda^2}{2\,\Delta_0}\int_0^bdx\,\expval{\left|\left(\partial_xh(x,t)\right)^2\right|^2}_{p_\text{s}[h]}\\
&=\frac{8\pi^4\lambda^2}{\Delta_0\,b^3}\sum_{k\in\mathcal{R}}\sum_{l,n\in\mathcal{R}_k\setminus\{0,k\}}l(k-l)n(k-n)\expval{h_l(t)h_{k-l}(t)h_{-n}(t)h_{n-k}(t)}_{p_\text{s}[h]}\\
&=\frac{\Delta_0\,\lambda^2}{2\,b\,\nu^2}\left[\Lambda^2+\frac{3\Lambda^2-\Lambda}{2}\right]=\frac{\Delta_0\,\lambda^2}{4\,b\,\nu^2}\left[5\,\Lambda^2-\Lambda\right],
\end{split}\label{eq:ExactSSSigma}
\end{align}
where $\mathcal{R}_k\equiv[\max(-\Lambda,-\Lambda+k),\min(\Lambda,\Lambda+k)]$ (see \cite{NiggemannSeifert2020}). Again, a comparison of \eqref{eq:ExactSSSigma} with the corresponding result from \cite{NiggemannSeifert2020} shows that the lowest order perturbational approximation is also exact for the case of the entropy production rate $\sigma$ in $1d$.\\
Thus, by using \eqref{eq:PsOfHNormalized}, we can calculate for the $(1+1)$ dimensional KPZ equation the exact expressions for the stationary current $J$ (see \eqref{eq:ExactSSCurrent}) and the entropy production rate $\sigma$ (see \eqref{eq:ExactSSSigma}) for arbitrary values of the coupling parameter. This implies that two of the three constituents of the TUR product $\mathcal{Q}$ are known exactly. Hence, we state the intermediate result 
\begin{equation}
\mathcal{Q}=\frac{\stot}{\expval{\Psi(t)}^2}\,\text{var}[\Psi(t)]=\left(5-\frac{1}{\Lambda}\right)\,\frac{\text{var}[\Psi(t)]}{\Delta_0\,b\,t}.\label{eq:KPZTURIntermediate}
\end{equation}
In the following sections we present two different approaches to obtain results for $\text{var}[\Psi(t)]$ in order to complement \eqref{eq:KPZTURIntermediate}.

\section{Scaling Behavior of $\text{var}[\Psi(t)]$}\label{sec:ScalingVar}

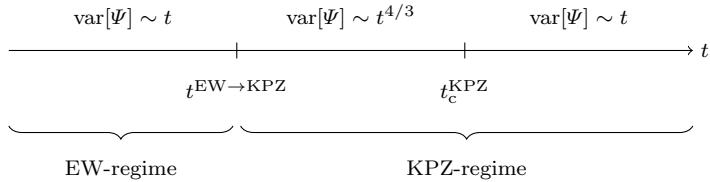
\begin{figure}[tbhp]
\centering
\begin{tikzpicture}
\draw[->] (0,0) -- (9,0) node[right] {$t$};
\draw (3 cm, 3pt) -- (3 cm, -3pt) node[align=center, below=5pt](a) {$t^{\text{EW}\to\text{KPZ}}$};
\draw (6 cm, 3pt) -- (6 cm, -3pt) node[align=center, below=5pt](b) {$t^\text{KPZ}_\text{c}$};
\node[align=center, above=5pt] at (1.5, 0) {$\text{var}[\Psi]\sim t$};
\node[align=center, above=5pt] at (4.5, 0) {$\text{var}[\Psi]\sim t^{4/3}$};
\node[align=center, above=5pt] at (7.5, 0) {$\text{var}[\Psi]\sim t$};
\draw[decorate, decoration = {brace, amplitude = 5pt, mirror}] (0, -1) -- (2.95, -1) node[midway, below=10pt] {$\text{EW-regime}$};
\draw[decorate, decoration = {brace, amplitude = 5pt, mirror}] (3.05, -1) -- (9, -1) node[midway, below=10pt] {$\text{KPZ-regime}$};
\end{tikzpicture}
\caption{Schematic illustration of the scaling behavior of $\text{var}[\Psi(t)]$ for the \lq normal\rq{} ordering of time-scales, i.e., $\tcorr>\tcross$, for a finite KPZ system (see, e.g., \cite{KrugReview1997}).}
\label{fig:ScalingVarianceSchematic}
\end{figure}
In contrast to the results in the previous section, which hold for any choice of system parameters, the variance of $\Psi(t)$ changes its behavior depending on the strength of the coupling parameter from \eqref{eq:DefLambdaEff}. To illustrate this in more detail, it is instructive to have a look at the time-scales at which the changes in the variance occur. In the case of a large coupling parameter, these time-scales are the EW to KPZ crossover time $\tcross$, given by \cite{KrugReview1997}
\begin{equation}
\tcross\approx252\,\nu^5\,\Delta_0^{-2}\,\lambda^{-4},\label{eq:EWtoKPZCrossoverTime}
\end{equation}
and the KPZ correlation time $\tcorr$, given by \cite{KrugReview1997}
\begin{equation}
\tcorr\approx\left(\frac{2\,(0.21)^3\,\nu}{\Delta_0}\right)^{1/2}\,\lambda^{-1}\,b^{3/2}.\label{eq:KPZCorrelationTime}
\end{equation}
In Fig. \ref{fig:ScalingVarianceSchematic}, we show schematically the behavior of the variance of $\Psi(t)$ if $\tcorr>\tcross$. For times $t<\tcross$ the system is in the so-called Edwards-Wilkinson regime, characterized by the critical exponent $z=2$ of the linear theory. In this scaling regime, the variance of $\Psi(t)$ is expected to scale linearly in time $t$ \cite{KrugReview1997}. For times in the range $\tcross<t<\tcorr$, the system is in its transient regime. This regime belongs to the KPZ scaling-regime, characterized by the KPZ critical exponent $z=3/2$. While in the transient regime, the variance is predicted via scaling arguments to scale with $t^{4/3}$, i.e., it displays super-diffusive behavior \cite{KrugReview1997}. For times $t>\tcorr$ the system enters the KPZ stationary regime, where the variance is again expected to scale linearly in time $t$. However, due to the super-diffusive behavior in the transient regime, the proportionality factor is larger in the stationary KPZ regime than in the EW scaling regime \cite{KrugReview1997}. In the following we will refer to the above described behavior as the behavior for the \lq normal\rq{} ordering of time-scales, namely $\tcorr>\tcross$.\\
Before we discuss the case of $\tcross>\tcorr$ let us reformulate the two time-scales in \eqref{eq:EWtoKPZCrossoverTime} and \eqref{eq:KPZCorrelationTime} by expressing both in terms of the effective coupling parameter $\leff$ from \eqref{eq:DefLambdaEff}. To this end we introduce a dimensionless time $t_s=t/T$ with the diffusive time scale $T=b^2/\nu$. This yields for the EW to KPZ crossover time 
\begin{equation}
\tcross_s\approx252\frac{\nu^6}{\Delta_0^2\,b^2}\,\frac{1}{\lambda^4}=\frac{252}{\leff^4},\label{eq:EWtoKPZTimeDimLess}
\end{equation}
and for the KPZ correlation time
\begin{equation}
t^{\text{KPZ}}_{\text{c},s}\approx\sqrt{2\,(0.21)^3}\left(\frac{\nu^3}{\Delta_0\,b}\right)^{1/2}\,\frac{1}{\lambda}=\frac{\sqrt{2\,(0.21)^3}}{\leff}.\label{eq:KPZCorrTimeDimLess}
\end{equation}
The form of \eqref{eq:EWtoKPZTimeDimLess} and \eqref{eq:KPZCorrTimeDimLess} indicates the existence of a critical effective coupling parameter $\leff^\text{c}$ below which the \lq normal\rq{} ordering of time-scales breaks down, i.e., $\tcross_s>t^{\text{KPZ}}_{\text{c},s}$. One may think of this as shrinking the transient regime in Fig. \ref{fig:ScalingVarianceSchematic} to zero, and thus, equating \eqref{eq:EWtoKPZTimeDimLess} with \eqref{eq:KPZCorrTimeDimLess} and solving for $\leff$ yields
\begin{equation}
\leff^\text{c}\approx12.28.\label{eq:DefCriticalLeff}
\end{equation}
In dependence of the critical effective coupling parameter we have
\begin{align}
\begin{split}
\tcross_s<t^{\text{KPZ}}_{\text{c},s}\quad&\text{for}\quad\leff>\leff^\text{c},\\
\tcross_s>t^{\text{KPZ}}_{\text{c},s}\quad&\text{for}\quad\leff<\leff^\text{c}.
\end{split}
\label{eq:TimeScalesOfLeff}
\end{align}
Hence, the behavior of $\text{var}[\Psi(t)]$ sketched in Fig. \ref{fig:ScalingVarianceSchematic} is valid as long as $\leff>\leff^\text{c}$.\\
We now turn to the behavior of the variance of $\Psi(t)$ for $\leff<\leff^\text{c}$. In this case we have $\tcross_s>t^{\text{KPZ}}_{\text{c},s}$, which is physically not sensible as this implies that the system would have to become stationary in the KPZ regime before even crossing over from the EW to the KPZ regime. This situation is resolved by taking the EW correlation time $\tcEW$ into account. It is given by \cite{KrugReview1997}
\begin{equation}
\tcEW=\frac{\pi\,b^2}{288\,\nu}\qquad\Leftrightarrow\qquad t_{\text{c},s}^\text{EW}=\frac{\pi}{288}.\label{eq:EWCorrTime}
\end{equation}
As can easily be seen, $\leff<\leff^\text{c}$ implies that $t_{\text{c},s}^\text{EW}<\tcross_s$, hence the system becomes stationary in the EW scaling regime. Therefore, if $\leff\ll\leff^\text{c}$, its dynamical behavior will be governed for all times $t$ by the critical exponent $z=2$ of the linear theory. For $\leff\uparrow\leff^\text{c}$ the behavior of the variance will change from the one in the linear theory to the one predicted for the KPZ equation and should be accessible by a perturbation expansion in $\leff$ up to $\leff\approx\leff^\text{c}$. Note, that when we state $\leff$ \lq small\rq{}, we mean $\leff<\leff^\text{c}$.

\section{Perturbation Expansion}\label{sec:PerturbationExpansion}

As stated in the above section, for values of $\leff\lessapprox\leff^\text{c}$ we expect to obtain the correct behavior of the variance of $\Psi(t)$ via a perturbation expansion in $\leff$. As the analysis follows the one in \cite{NiggemannSeifert2020}, we will be brief here and focus on the results instead of the technical details. Note, that in this section we use the scaled version of the KPZ equation, which is obtained by the introduction of scaled variables according to
\begin{equation}
x\to\frac{x}{b}\,,\qquad t\to\frac{t}{T}\,,\qquad h\to\frac{h}{H}\,,\qquad\eta\to\frac{\eta}{N}\,,\label{eq:DefScaledVariables}
\end{equation}
with $T=b^2/\nu$, $H=\sqrt{\Delta_0b/\nu}$ and $N=\sqrt{\Delta_0\nu/b^3}$ \cite{NiggemannSeifert2020}. This significantly simplifies the perturbation expansion. In the scaled variables the perturbative ansatz reads
\begin{equation}
h(x,t)=\sum_{k\in\mathcal{R}}\left[h_k^{(0)}(t)+\leff\,h_k^{(1)}(t)+\leff^2\,h_k^{(2)}(t)+O(\leff^3)\right]\,e^{2\pi ikx},\label{eq:DefPerturbationExpansionH}
\end{equation}
where the Fourier coefficients $h_k^{(i)}(t)$ are given in \cite{NiggemannSeifert2020}. In the following we will use \eqref{eq:DefPerturbationExpansionH} to evaluate 
\begin{equation}
\text{var}[\Psi(t)]=\int_0^td\tp\int_0^td\tpp\expval{\dot{\Psi}(\tp)\dot{\Psi}(\tpp)},\label{eq:DefVarOfPsi}
\end{equation}
where
\begin{equation}
\dot{\Psi}(t)=\frac{\leff}{2}\int_0^1dx\left(\partial_xh(x,t)\right)^2+\int_0^1dx\,\eta(x,t).\label{eq:DefPsiDot}
\end{equation}
Here, the dimensionless form of \eqref{eq:KPZEq}, given in \cite{NiggemannSeifert2020}, is integrated with respect to the spatial variable to obtain \eqref{eq:DefPsiDot}. We use \eqref{eq:DefPsiDot} to calculate the two-time correlation function
\begin{align}
\begin{split}
\expval{\dot{\Psi}(\tp)\,\dot{\Psi}(\tpp)}&=\frac{\leff^2}{4}\int_0^1dx\int_0^1dy\,\expval{\left(\partial_xh(x,\tp)\right)^2\left(\partial_yh(y,\tpp)\right)^2}+\delta(\tp-\tpp)\\
&\equiv\frac{\leff^2}{4}\,\mathcal{J}(\tp,\tpp)+\delta(\tp-\tpp),
\end{split}\label{eq:DefExpValPsiDotPsiDot}
\end{align}
where
\begin{equation}
\mathcal{J}(\tp,\tpp)=(2\pi)^4\sum_{k,l\in\mathcal{R}\setminus\{0\}}k^2\,l^2\,\expval{h_k(\tp)h_{-k}(\tp)h_l(\tpp)h_{-l}(\tpp)}.\label{eq:DefAuxJ}
\end{equation}
In principle any correlation of the Fourier coefficients $h_k$ can eventually be expressed by correlations of $h_k^{(0)}$ from \eqref{eq:DefPerturbationExpansionH}, which depend linearly on the Gaussian noise $\eta$ \cite{NiggemannSeifert2020} and thus allow for the application of Wick's theorem. In practice, however, this results in a quickly growing complexity of the calculation for higher order approximations in $\leff$. A possible circumvention of this issue is the physical assumption of so-called quasi-normality. This assumption has been successfully used in turbulence theory \cite{McCombBook,Sagaut2018} and has been adopted in \cite{Krug1991} for the height field $h(x,t)$ of the KPZ equation. The quasi-normality hypothesis states that all even moments of $h$ are assumed to behave like they were normally distributed and thus Wick's theorem may directly be applied to \eqref{eq:DefAuxJ}. At least for large times $\tp$, $\tpp$ the assumption is supported by the fact that $h(x,t)$ is exactly Gaussian distributed in the NESS (see \autoref{sec:ExactResults}).\\
Hence, after applying Wick's theorem to \eqref{eq:DefAuxJ}, we have
\begin{align}
\begin{split}
&\expval{\dot{\Psi}(\tp)\,\dot{\Psi}(\tpp)}-\expval{\dot{\Psi}(\tp)}\expval{\dot{\Psi}(\tpp)}\\
&=\delta(\tp-\tpp)+(2\pi)^4\frac{\leff^2}{4}\sum_{k,l\in\mathcal{R}\setminus\{0\}}k^2\,l^2\\
&\times\left[\expval{h_k(\tp)h_l(\tpp)}\expval{h_{-k}(\tp)h_{-l}(\tpp)}+\expval{h_k(\tp)h_{-l}(\tpp)}\expval{h_{-k}(\tp)h_l(\tpp)}\right].
\end{split}\label{eq:VarPsiDot}
\end{align}
Replacing the $h_j$'s in \eqref{eq:VarPsiDot} with the expansion from \eqref{eq:DefPerturbationExpansionH}, integrating twice over time and following the same steps as in \cite{NiggemannSeifert2020}, one obtains for $t\gg1$
\begin{align}
\begin{split}
\text{var}[\Psi(t)]&\simeq t\,\left[1+\frac{\leff^2}{32\pi^2}\sum_{k\in\mathcal{R}\setminus\{0\}}\frac{1}{k^2}\right.\\
&-\left.\frac{\leff^4}{256\pi^4}\sum_{k\in\mathcal{R}\setminus\{0\}}\sum_{m\in\mathcal{R}_k\setminus\{0,k\}}\frac{m}{k^3(k^2+(k-m)^2+m^2)}+O(\leff^6)\right]\\
&\equiv t\,\left[1+\frac{\leff^2}{32\pi^2}\mathcal{S}_1(\Lambda)-\frac{\leff^4}{256\pi^4}\mathcal{S}_2(\Lambda)+O(\leff^6)\right],
\end{split}\label{eq:VarOfPsiRes}
\end{align}
which is calculated to one order higher than in \cite{NiggemannSeifert2020} and $\mathcal{S}_{1,2}$ are simply abbreviations for the respective sums in the first line. Next we will evaluate $\mathcal{S}_{1,2}$ analytically in the limit of large $\Lambda$. For $\mathcal{S}_1$ we find
\begin{equation}
\mathcal{S}_1(\Lambda)=2\mathcal{H}_\Lambda^{(2)}\quad\stackrel{\Lambda\gg1}{\longrightarrow}\quad 2\,\zeta(2),\label{eq:EvalSum1}
\end{equation}
where $\mathcal{H}_\Lambda^{(n)}=\sum_{k=1}^\Lambda1/k^n$ is the so-called generalized harmonic number of order $n$ and $\zeta$ the Riemann-Zeta function. For $\mathcal{S}_2$ we find after some straightforward algebraic manipulation
\begin{align}
\begin{split}
\mathcal{S}_2(\Lambda)&=\sum_{k\in\mathcal{R}\setminus\{0\}}\left[\sum_{m\in\mathcal{R}_k}\frac{m}{k^3(k^2+(k-m)^2+m^2)}-\frac{1}{2\,k^4}\right]\\
&=2\sum_{k=1}^\Lambda\frac{1}{k^3}\sum_{m=-\Lambda+k}^\Lambda\frac{m}{k^2+(k-m)^2+m^2}-\mathcal{H}_\Lambda^{(4)}.
\end{split}\label{eq:EvalSum2Step1}
\end{align}
The inner sum over $m$ in the second line in \eqref{eq:EvalSum2Step1} may be approximated by the model $a-b\,k/\Lambda$, with $a$, $b$ free fit-parameters, which we estimated as $a\approx0.9065$, $b\approx0.6045$, and thus
\begin{equation}
\sum_{m=-\Lambda+k}^\Lambda\frac{m}{k^2+(k-m)^2+m^2}\approx0.9065-\frac{0.6045}{\Lambda}\,k.\label{eq:EvalSum2Step2}
\end{equation}
Inserting \eqref{eq:EvalSum2Step2} into \eqref{eq:EvalSum2Step1} and taking $\Lambda\gg1$ leads to
\begin{equation}
2\sum_{k=1}^\Lambda\frac{1}{k^3}\sum_{m=-\Lambda+k}^\Lambda\frac{m}{k^2+(k-m)^2+m^2}-\mathcal{H}_\Lambda^{(4)}\quad\stackrel{\Lambda\gg1}{\longrightarrow}\quad1.813\,\zeta(3)-\zeta(4).\label{eq:EvalSum2Final}
\end{equation}
Thus in the case of large $\Lambda$ we have the following asymptotic behavior ($t\gg1$) of the variance of $\Psi$,
\begin{equation}
\text{var}[\Psi(t)]\simeq t\,\left[1+\frac{\leff^2}{16\pi^2}\,\zeta(2)-\frac{\leff^4}{256\pi^4}\left(1.813\,\zeta(3)-\zeta(4)\right)+O(\leff^6)\right],\label{eq:VarPerturbationResultDimLess}
\end{equation}
or, in terms of the rescaled, dimensional variables
\begin{equation}
\text{var}[\Psi(t)]\simeq\Delta_0\,b\,t\left[1+\frac{\Delta_0b\lambda^2}{16\pi^2\nu^3}\zeta(2)-\frac{\Delta_0^2b^2\lambda^4}{256\pi^4\nu^6}\left(1.813\,\zeta(3)-\zeta(4)\right)+O(\leff^6)\right].\label{eq:VarPerturbationResultDim}
\end{equation}
We expect the approximations in \eqref{eq:VarPerturbationResultDimLess} and \eqref{eq:VarPerturbationResultDim} to yield sound results for $\leff\lessapprox\leff^\text{c}$ from \eqref{eq:DefCriticalLeff}. In \autoref{sec:CompNumSim} this will be checked by comparison with numerical simulations in the according parameter regime.\\
In the next section, we will focus on obtaining the variance of $\Psi(t)$ for large values of $\leff$ via a DRG approach.

\section{Dynamic Renormalization Group Calculation}\label{sec:RGCalculation}

\subsection{The $1D$ KPZ-Burgers Equation and $\text{var}[\Psi(t)]$}\label{subsec:RG_1DBurgersVar}

In this section we use the equivalence of the $1d$ KPZ equation to the stochastic Burgers equation (see e.g. \cite{Burgers1994}), given by the transformation $u(x,t)=-\partial_xh(x,t)$, with $u(x,t)$ the velocity field of the Burgers equation
\begin{equation}
\partial_tu(x,t)+\frac{\lambda}{2}\partial_xu^2(x,t)=\nu\partial_x^2u(x,t)+f(x,t),\label{eq:DefBurgersEq}
\end{equation} 
where $f(x,t)=-\partial_x\eta(x,t)$. In terms of the Burgers velocity field $u(x,t)$ the expression for $\dot{\Psi}(t)$ from \eqref{eq:DefPsiDot} reads
\begin{equation}
\dot{\Psi}(t)=\frac{\lambda}{2}\int_0^bdx\,u^2(x,t)+\int_0^bdx\,\eta(x,t).\label{eq:DefVarOfU}
\end{equation}
In principle, the derivation of the expression for the variance of $\Psi(t)$ is analogous to the one shown in \autoref{sec:PerturbationExpansion}. However, here we will use the continuous Fourier transform instead of the discrete Fourier series as above, since a continuous wavenumber spectrum is needed for implementing the DRG scheme. In particular, we define
\begin{equation}
u(q,\omega)=\int dx\int dt\,u(x,t)\,e^{-i(qx-\omega t)}\,,\quad u(x,t)=\int \frac{dq}{2\pi}\int \frac{d\omega}{2\pi}\,u(q,\omega)\,e^{i(qx-\omega t)},\label{eq:DefFourierTransform}
\end{equation}
as the forward and backward Fourier transform of the velocity field $u(x,t)$, respectively. To apply \eqref{eq:DefFourierTransform} to \eqref{eq:DefVarOfU}, we use the $b$-periodicity of $u(x,t)$ due to the periodic boundary conditions in \eqref{eq:KPZEq}. In particular we have
\begin{align}
\begin{split}
\int_0^bdx\,u^2(x,t)&=\int_{-b/2}^{b/2}dx\,u^2(x,t)\\
&\approx\int_{-\infty}^\infty dx\,u^2(x,t)=\frac{1}{2\,\pi}\int_{-\infty}^\infty dq\,u(q,t)\,u(-q,t),\label{eq:ApplicationFT}
\end{split}
\end{align}
where the second step holds for $b\gg1$ and in the last step we used the partial Fourier transform \eqref{eq:DefFourierTransform} in the spatial variable $x$. We thus obtain
\begin{align}
\expval{\Psi(t)}&\approx\frac{\lambda}{2}\int_0^td\tp\,\int\frac{dq}{2\pi}\expval{u(q,\tp)\,u(-q,\tp)},\label{eq:ExpValPsiOfU}\\
\begin{split}
\expval{\Psi^2(t)}&\approx\expval{\Psi(t)}^2+\Delta_0\,b\,t+\frac{\lambda^2}{4}\int_0^td\tp\int_0^td\tpp\int\frac{dq}{2\pi}\int\frac{d\qp}{2\pi}\\
&\times\left[\expval{u(q,\tp)u(\qp,\tpp)}\expval{u(-q,\tp)u(-\qp,\tpp)}\right.\\
&\qquad+\left.\expval{u(q,\tp)u(-\qp,\tpp)}\expval{u(-q,\tp)u(\qp,\tpp)}\right],
\end{split}\label{eq:ExpValPsiSquaredOfU}
\end{align}
and therefore, similar to \eqref{eq:DefVarOfPsi},
\begin{align}
\begin{split}
\text{var}[\Psi(t)]&\approx\Delta_0\,b\,t+\frac{\lambda^2}{4}\int_0^td\tp\int_0^td\tpp\int\frac{dq}{2\pi}\int\frac{d\qp}{2\pi}\\
&\times\left[\expval{u(q,\tp)u(\qp,\tpp)}\expval{u(-q,\tp)u(-\qp,\tpp)}\right.\\
&\qquad+\left.\expval{u(q,\tp)u(-\qp,\tpp)}\expval{u(-q,\tp)u(\qp,\tpp)}\right].
\end{split}\label{eq:DefVarPsiOfU}
\end{align}
The expressions in \eqref{eq:ExpValPsiSquaredOfU} and \eqref{eq:DefVarPsiOfU} again rely on the quasinormality hypothesis \cite{Krug1991,McCombBook}. 

\subsection{Two-Point Correlation Function via DRG}\label{subsec:RG_CorrelationFunction}

Instead of calculating the two-point correlation functions in \eqref{eq:DefVarPsiOfU} perturbatively as in \autoref{sec:PerturbationExpansion}, we here use the DRG method described in e.g. \cite{FNS1977,YakhotShe1988}, where we have noise correlations corresponding to Gaussian white noise for the KPZ equation, i.e.,
\begin{equation}
\expval{f(q,\omega)f(\qp,\op)}=-(2\pi)^2\,\Delta_0\,q\,\qp\,\delta(q+\qp)\,\delta(\omega+\op),\label{eq:NoiseCorrelationFourierBurgers}
\end{equation}
($y=-2$ in \cite{FNS1977,YakhotShe1988}). The starting point of the DRG procedure is the Fourier-space representation of \eqref{eq:DefBurgersEq}, namely
\begin{equation}
u(q,\omega)=G_0(q,\omega)f(q,\omega)-iq\frac{\lambda}{2}G_0(q,\omega)\int\frac{d\qp}{2\pi}\int\frac{d\op}{2\pi}u(\qp,\op)u(q-\qp,\omega-\op),\label{eq:DefUOfQOmega}
\end{equation}
where we define the bare propagator
\begin{equation}
G_0(q,\omega)\equiv\frac{1}{-i\,\omega+\nu\,q^2}.\label{eq:DefBarePropagator}
\end{equation}
The next step will be to split the velocity field in \eqref{eq:DefUOfQOmega} into large-wavenumber modes, $u^>$, and small-wavenumber modes, $u^<$, where it holds that (see e.g. \cite{McCombBook,FNS1977})
\begin{equation}
u(q,\omega)=\begin{cases} u^<(q,\omega) \quad\text{for}\quad 0<q<\Lo\,e^{-l}, \\ u^>(q,\omega) \quad\text{for}\quad \Lo\,e^{-l}<q<\Lo, \end{cases}
\end{equation}
with $l$ the renormalization parameter and $\Lo$ an ultraviolet wavenumber cutoff. An analogous splitting applies to the noise $f(q,\omega)$ as well. Averaging the ensuing equations with respect to the noise history of the $f^>$-modes and integrating out the contributions of the large-wavenumber modes $u^>$ yields corrections to the terms of the small-wavenumber modes $u^<$. As these steps are well known and explained in detail in e.g. \cite{McCombBook,FNS1977}, we will simply state the results, which are the renormalization equations for $\nu$ and $\Delta_0$,
\begin{align}
\nu_\text{R}=\nu\left[1+\frac{\lambda^2\,\Delta_0}{8\pi\,\nu^3}\frac{e^l-1}{\Lo}\right],\label{eq:RGEqNu}\\
\Delta_{0,\text{R}}=\Delta_0\left[1+\frac{\lambda^2\,\Delta_0}{8\pi\,\nu^3}\frac{e^l-1}{\Lo}\right],\label{eq:RGEqDelta0}
\end{align}
obtained after one elimination step. This mode elimination process is iterated using infinitesimally small wavenumber increments ($l\to dl$) which causes parameter changes $d\nu$ and $d\Delta_0$. One thus arrives at differential equations for $\nu(l)$ and $\Delta_0(l)$, given respectively by
\begin{align}
\frac{\text{d}\,\nu(l)}{\text{d}l}&=\nu(l)\,\frac{\overline{\lambda}^2}{8\,\pi},\label{eq:DiffRGEqNu}\\
\frac{\text{d}\,\Delta_0(l)}{\text{d}l}&=\Delta_0(l)\,\frac{\overline{\lambda}^2}{8\,\pi},\label{eq:DiffRGEqDelta0}
\end{align}
where
\begin{equation}
\overline{\lambda}\equiv\lambda\left(\frac{\Delta_0(l)}{\nu^3(l)\Lp(l)}\right)^{1/2},\label{eq:DefLambdaBar}
\end{equation}
with $\Lp(l)\equiv \Lo e^{-l}$ (see e.g. \cite{YakhotShe1988}). $\overline{\lambda}(l)$ denotes the renormalized coupling constant characteristic for the eliminated modes. Up to a constant numerical prefactor, $\overline{\lambda}(0)$ equals $\leff$ from \eqref{eq:DefLambdaEff}. At this point we adopt a DRG scheme introduced in \cite{YakhotOrszag1986_1,YakhotOrszag1986_2} and analyzed in \cite{SmithWoodruff1998}, which has been recently applied in \cite{Singha2014,Rodriguez2020}. It implies that the next step of the scheme consists in solving \eqref{eq:DiffRGEqNu} and \eqref{eq:DiffRGEqDelta0} for $\nu$ and $\Delta_0$ explicitly, making their scale dependence transparent. It follows directly that,
\begin{equation}
\frac{\Delta_0(l)}{\nu(l)}=\frac{\Delta_0}{\nu}=\text{const.}\quad \forall\, l>0\,,\label{eq:ConstRatioNuDelta0}
\end{equation}
with $\nu$, $\Delta_0$ and $\Lo$ the unrenormalized parameters from \eqref{eq:DefBurgersEq} and \eqref{eq:NoiseCorrelationFourierBurgers}. The finding in \eqref{eq:ConstRatioNuDelta0} reflects the fluctuation-dissipation theorem, known to hold for the $1d$ Burgers-KPZ system (see e.g. \cite{KrugReview1997,FNS1977,TaeuberFrey1994}). Using \eqref{eq:DefLambdaBar} and \eqref{eq:ConstRatioNuDelta0}, the integration of \eqref{eq:DiffRGEqNu} yields
\begin{equation}
\nu(l)=\nu\left(1+\frac{\lambda^2\Delta_0}{4\pi\nu^3}\frac{e^l-1}{\Lo}\right)^{1/2}.\label{eq:SolutionNuOfL}
\end{equation}
As a last step we make the common identification $|q|=\Lo e^{-l}$ (see e.g. \cite{McCombBook,FNS1977,YakhotShe1988,YakhotOrszag1986_1,YakhotOrszag1986_2,Singha2014,Rodriguez2020}) and obtain asymptotically for large values of $l$ (i.e. $|q|\ll1$)
\begin{align}
\nu(q)&\simeq\frac{\lambda}{2\,\sqrt{\pi}}\,\left(\frac{\Delta_0}{\nu}\right)^{1/2}\,|q|^{-1/2},\label{eq:SolutionNuOfQ}\\
\Delta_0(q)&=\frac{\Delta_0}{\nu}\,\nu(q)\simeq\frac{\lambda}{2\,\sqrt{\pi}}\,\left(\frac{\Delta_0}{\nu}\right)^{3/2}\,|q|^{-1/2}.\label{eq:SolutionDelta0OfQ}
\end{align}
Equivalently, $\bar{\lambda}(l)$ converges for $l\to\infty$ (i.e. after all large wavenumber modes are eliminated) to a finite stable fixed point, the KPZ fixed point of the RG-flow. This fixed point is associated with the dynamical scaling exponent $z=3/2$.\\
According to \cite{YakhotShe1988}, the expressions from \eqref{eq:SolutionNuOfQ} and \eqref{eq:SolutionDelta0OfQ} allow for the introduction of a renormalized effective propagator
\begin{equation}
G(q,\omega)=\frac{1}{-i\,\omega+\nu(q)\,q^2},\label{eq:DefRGPropagator}
\end{equation}
and a renormalized effective noise with
\begin{equation}
\expval{f(q,\omega)\,f(\qp,\op)}=(2\pi)^2\,\Delta_0(q)\,q^2\,\delta(q+\qp)\,\delta(\omega+\op),\label{eq:DefRGNoiseCorr}
\end{equation}
such that the nonlinear equation from \eqref{eq:DefUOfQOmega} may be replaced by an effective linear Langevin equation
\begin{equation}
u(q,\omega)\simeq G(q,\omega)\,f(q,\omega).\label{eq:UofQOmegaAfterRG}
\end{equation}
Note, that in \eqref{eq:UofQOmegaAfterRG}, as opposed to \eqref{eq:DefUOfQOmega}, the right hand side now depends on $\nu(q)$ and $\Delta_0(q)$ from \eqref{eq:SolutionNuOfQ} and \eqref{eq:SolutionDelta0OfQ}, respectively. A justification of this step is given in \cite{YakhotShe1988,YakhotOrszag1986_1,YakhotOrszag1986_2,SmithWoodruff1998} via the so-called $\epsilon$-expansion. In the present case, a further justification may be given by the fact that for large times the fluctuations of $h(x,t)$ become Gaussian distributed. This indicates that their dynamics may be described by a linear Langevin equation as in \eqref{eq:UofQOmegaAfterRG}. An analogous conclusion has been drawn for a slightly different setting in \cite{Rodriguez2020}.\\
Using \eqref{eq:DefRGPropagator} and \eqref{eq:DefRGNoiseCorr}, we give an explicit approximation for the two-point correlation $\expval{u(q,\omega)u(\qp,\op)}$,
\begin{align}
\begin{split}
\expval{u(q,\omega)\,u(\qp,\op)}&\simeq G(q,\omega)\,G(\qp,\op)\,\expval{f(q,\omega)\,f(\qp,\op)}\\
&=(2\pi)^2\,q^2\,\Delta_0(q)\,G(q,\omega)\,G(\qp,\op)\,\delta(q+\qp)\,\delta(\omega+\op)\\
&\equiv(2\pi)^2\,C(q,\omega)\,\delta(q+\qp)\,\delta(\omega+\op).
\end{split}\label{eq:DefCofQOmega}
\end{align}
Here we have introduced the correlation function $C(q,\omega)$ (see e.g. \cite{YakhotShe1988}) according to
\begin{equation}
C(q,\omega)\equiv q^2\,\Delta_0(q)\,G(q,\omega)\,G(-q,-\omega)=q^2\,\Delta_0(q)\,\left|G(q,\omega)\right|^2,\label{eq:DefCofQOmegaExplicit}
\end{equation}
with $\Delta_0(q)$ from \eqref{eq:SolutionDelta0OfQ} and $G(q,\omega)$ from \eqref{eq:DefRGPropagator}. Inserting the explicit expressions form \eqref{eq:SolutionNuOfQ} and \eqref{eq:SolutionDelta0OfQ} into \eqref{eq:DefCofQOmegaExplicit} we arrive at (see also \cite{YakhotShe1988} for $\lambda=1$)
\begin{equation}
C(q,\omega)\approx\left(\frac{\Delta_0}{\nu}\right)^{1/2}\frac{2\pi^{1/2}}{\lambda}\,\frac{|q|^{-3/2}}{1+\left(\left(\frac{\nu}{\Delta_0}\right)^{1/2}\frac{2\pi^{1/2}}{\lambda}\,\frac{\omega}{|q|^{3/2}}\right)^2}.\label{eq:CofQOmegaScaling}
\end{equation}
Obviously, \eqref{eq:CofQOmegaScaling} is in accordance with the well known scaling result for the correlation function of the Burgers equation in one spatial dimension (see e.g. \cite{FNS1977,YakhotShe1988}),
\begin{equation}
C(q,\omega)\sim |q|^{-3/2}\Phi\left(\frac{\omega}{|q|^{3/2}}\right), \label{eq:DefScalingBurgersC}
\end{equation}
with $\Phi$ as a universal scaling function.

\subsection{DRG Results for $\text{var}[\Psi(t)]$}\label{subsec:RG_ResultsVar}

Performing a Fourier backward transformation in frequency on both sides of \eqref{eq:DefCofQOmega} and inserting for $C(q,\omega)$ the expression from \eqref{eq:CofQOmegaScaling} leads to
\begin{align}
\begin{split}
\expval{u(q,t)\,u(\qp,\tp)}&=2\pi\,C(q,t-\tp)\,\delta(q+\qp)\\
&\approx2\pi\,\frac{\Delta_0}{2\,\nu}\,\exp\left[-\left(\frac{\Delta_0}{2\,\nu}\right)^{1/2}\frac{\lambda}{(2\pi)^{1/2}}\,|q|^{3/2}\,|t-\tp|\right]\,\delta(q+\qp),\label{eq:DefTwoPointCorrelationUofQT}
\end{split}
\end{align}
as the approximate two-point correlation of $u=-\partial_xh$ in wavenumber space. With \eqref{eq:DefTwoPointCorrelationUofQT} we now have the necessary means to calculate the product of two-point correlation functions in \eqref{eq:DefVarPsiOfU}. In particular,
\begin{align}
\begin{split}
&\int\frac{dq}{2\pi}\int\frac{d\qp}{2\pi}\,\expval{u(q,\tp)\,u(\qp,\tpp)}\,\expval{u(-q,\tp)\,u(-\qp,\tpp)}\\
&=\int\frac{dq}{2\pi}\int\frac{d\qp}{2\pi}\,C(q,\tp-\tpp)\,C(-\qp,\tp-\tpp)\,\delta(q+\qp)\,\delta(-q-\qp)\\
&=\int dk\int dk^\prime\,C\left(\frac{2\pi}{b}\,k,\tp-\tpp\right)\,C\left(-\frac{2\pi}{b}\,k^\prime,\tp-\tpp\right)\,\delta(k+k^\prime)\\
&=\frac{b}{4\pi}\,\left(\frac{\Delta_0}{\nu}\right)^{5/3}\frac{\Gamma(5/3)\,\pi^{1/3}}{\lambda^{2/3}}\,\frac{1}{|\tp-\tpp|^{2/3}},
\end{split}\label{eq:ResultProductTwoPointCorrUofQT1}
\end{align}
where we substituted in the second step $q=2\pi k/b$ to attribute for the fact that we operate on a finite system-size $x\in[0,b]$, which implies an explicit length scale, and $\Gamma$ is the Euler-Gamma function. Analogously, the second term in \eqref{eq:DefVarPsiOfU} becomes
\begin{align}
\begin{split}
&\int\frac{dq}{2\pi}\int\frac{d\qp}{2\pi}\,\expval{u(q,\tp)\,u(-\qp,\tpp)}\,\expval{u(-q,\tp)\,u(\qp,\tpp)}\\
&=\frac{b}{4\pi}\,\left(\frac{\Delta_0}{\nu}\right)^{5/3}\frac{\Gamma(5/3)\,\pi^{1/3}}{\lambda^{2/3}}\,\frac{1}{|\tp-\tpp|^{2/3}}
\end{split}\label{eq:ResultProductTwoPointCorrUofQT2}
\end{align}
Inserting \eqref{eq:ResultProductTwoPointCorrUofQT1} and \eqref{eq:ResultProductTwoPointCorrUofQT2} into \eqref{eq:DefVarPsiOfU} yields
\begin{align}
\begin{split}
\text{var}[\Psi(t)]&\approx\Delta_0\,b\,t+\frac{\Gamma(5/3)}{8\,\pi^{2/3}}\,b\,\frac{\lambda^{4/3}\,\Delta_0^{5/3}}{\nu^{5/3}}\int_0^td\tp\int_0^td\tpp\,|\tp-\tpp|^{-2/3}\\
&=\Delta_0\,b\,t+\frac{3\,\Gamma(2/3)}{8\,\pi^{2/3}}\,b\,\lambda^{4/3}\,\left(\frac{\Delta_0}{\nu}\right)^{5/3}\,t^{4/3}.
\end{split}\label{eq:ResultVarPsiOfU}
\end{align}
Therefore in the long-time limit,
\begin{equation}
\text{var}[\Psi(t)]\simeq\frac{3\,\Gamma(2/3)}{8\,\pi^{2/3}}\,b\,\lambda^{4/3}\,\left(\frac{\Delta_0}{\nu}\right)^{5/3}\,t^{4/3},\label{eq:ResultVarPsiOfULongTimeLimit}
\end{equation}
which indicates super-diffusive behavior for the variance of $\Psi(t)$. This expression is in accordance with the scaling form predicted in \cite{KrugReview1997} for the transient KPZ regime. Moreover, the present DRG approach yields the amplitude factors as well. We use \eqref{eq:ResultVarPsiOfULongTimeLimit} in the time range $\tcross<t<\tcorr$ and check in the following for consistency with known results at the endpoints of this time interval. For times $t\geq\tcorr$, the variance of $\Psi(t)$ behaves as $\text{var}[\Psi(t)]=\mathcal{C}\,t$ (see Fig. \ref{fig:ScalingVarianceSchematic}), with $\mathcal{C}$ a parameter to be determined. Hence, we have the matching condition
\begin{equation}
\text{var}[\Psi(\tcorr)]\simeq\frac{3\,\Gamma(2/3)}{8\,\pi^{2/3}}\,b\,\lambda^{4/3}\,\left(\frac{\Delta_0}{\nu}\right)^{5/3}\,(\tcorr)^{4/3}\stackrel{!}{=}\mathcal{C}\,\tcorr.\label{eq:MatchingConditionAtCorrelationTime}
\end{equation}
Inserting $\tcorr$ from \eqref{eq:KPZCorrelationTime} into \eqref{eq:MatchingConditionAtCorrelationTime} and solving for $\mathcal{C}$ yields
\begin{equation}
\mathcal{C}\approx0.3444\,\left(\frac{\Delta_0}{2\,\nu}\right)^{3/2}\,\lambda\,b^{3/2}.\label{eq:ResultMatchingConstant1}
\end{equation}
This expression may be compared with a result in \cite{KrugReview1997} for the quantity $W_c^2$, which differs from $\text{var}[\Psi(t)]$ only by a factor of $1/b^2$, given by
\begin{equation}
W_c^2=c_0\,\left(\frac{\Delta_0}{2\,\nu}\right)^{3/2}\,\lambda\,b^{-1/2}\,t,\label{eq:KrugVar}
\end{equation} 
with $c_0$ a universal scaling amplitude. Apart from the prefactor $c_0$ the result in \eqref{eq:ResultMatchingConstant1} is the same as the one in \eqref{eq:KrugVar}, in particular with respect to the anomalous scaling in $b$. Regarding $c_0$, this was determined in \cite{Derrida1993} for the ASEP-process and then adopted in \cite{KrugReview1997} relying on the universality hypothesis. The exact value of $c_0$ reads
\begin{equation}
c_0=\frac{\sqrt{\pi}}{4}\approx0.443.\label{eq:DefC0ScalingAmplitude}
\end{equation} 
We thus deviate from the exact result for $c_0$ by roughly $20\%$, which we regard satisfactory for our consistency check. Moreover we note that our numerical simulations in \autoref{sec:CompNumSim} indicate that the numerical values resulting from the theoretical prediction of $\tcorr$ from \eqref{eq:KPZCorrelationTime} are too small by roughly a factor of $2$ see \autoref{subsec:NumComp_Var}. Taking this into account, the correspondence between the numerical values from \eqref{eq:ResultMatchingConstant1} and \eqref{eq:DefC0ScalingAmplitude} is improved significantly (see \eqref{eq:NumC0AmplitudeResult}).\\
At the left endpoint of the transient KPZ regime, i.e., at $t=\tcross$, consistency may be checked by comparing \eqref{eq:ResultVarPsiOfULongTimeLimit} with the perturbation expansion from \eqref{eq:VarPerturbationResultDim} for $\leff\approx\leff^\text{c}$. This makes sense, since on the one hand we know that $\tcorr\gtrapprox\tcross$ provided that $\leff\gtrapprox\leff^\text{c}$. On the other hand the expansion from \eqref{eq:VarPerturbationResultDim} is expected to be valid for $\leff\lessapprox\leff^\text{c}$. Thus, with \eqref{eq:ResultVarPsiOfULongTimeLimit} we get
\begin{align}
\begin{split}
\text{var}[\Psi(\tcross)]&\simeq\frac{3\,\Gamma(2/3)}{8\,\pi^{2/3}}\,b\,\lambda^{4/3}\,\left(\frac{\Delta_0}{\nu}\right)^{5/3}\,(\tcross)^{4/3}\\
&\approx1.4953\,\Delta_0\,b\,\tcross,
\end{split}\label{eq:MatchingConditionAtCrossoverTime}
\end{align}
whereas evaluating \eqref{eq:VarPerturbationResultDim} at $\leff=\leff^\text{c}\approx12.28$ from \eqref{eq:DefCriticalLeff} results in 
\begin{equation}
\text{var}[\Psi(\tcross)]\approx1.5704\,\Delta_0\,b\,\tcross.\label{eq:VarPerAtCrossoverTimeForLeffC}
\end{equation}
Hence, the respective results differ by just $5\%$. Taking into consideration that both results in \eqref{eq:VarPerturbationResultDim} and \eqref{eq:ResultVarPsiOfULongTimeLimit} are approximations, this seems to be a reasonable match.\\
To sum up, this section was devoted to the derivation of an analytical expression approximating $\text{var}[\Psi(t)]$ in the transient and steady state KPZ regime, respectively. Whereas the result for the latter is essentially known from \cite{KrugReview1997}, the result for the former seems to be new. We stress that all amplitude factors are determined by analytic calculations for a generic KPZ system, i.e., without invoking specific model problems of the KPZ universality class. Furthermore, our approximation from \eqref{eq:DefTwoPointCorrelationUofQT} for the two-point correlation of $u=-\partial_xh$ in wavenumber space may be of some interest in itself. This is since the exact scaling function found in \cite{Praehofer2004} for the $1d$ KPZ equation is given via the solution of certain differential equations (Painlev\'{e} II), which can be solved only by quite involved numerical methods. Especially, an exact analytic expression seems to be out of reach. 

\section{Thermodynamic Uncertainty Relation}\label{sec:TURFull}

Before we formulate the TUR for an arbitrary value of the coupling parameter, let us collect what we have derived for the variance of $\Psi(t)$ in the above sections. Consider first the parameter regime where $\leff<\leff^\text{c}$. Here we know from \eqref{eq:VarPerturbationResultDimLess} that for times $t>\tcEW$
\begin{equation}
\text{var}[\Psi(t)]\simeq\Delta_0\,b\,t\left[1+\frac{\leff^2}{16\pi^2}\,\zeta(2)-\frac{\leff^4}{256\pi^4}\left(1.813\,\zeta(3)-\zeta(4)\right)+O(\leff^6)\right],\label{eq:VarScalingLeffSmallerLeffC}
\end{equation}
whereas for $t\ll\tcEW$, $\text{var}[\Psi(t)]\simeq\Delta_0\,b\,t$ holds.\\
On the other hand for a parameter set with $\leff\gg\leff^\text{c}$ we have shown in \eqref{eq:ResultVarPsiOfU} that
\begin{align}
\begin{split}
\text{var}[\Psi(t)]\simeq\begin{cases}  
\Delta_0\,b\,t\,,\quad  &t\ll\tcross, \\ \\
\frac{3\,\Gamma(2/3)}{8\,\pi^{2/3}}\,b\,\lambda^{4/3}\,\left(\frac{\Delta_0}{\nu}\right)^{5/3}\,t^{4/3}\,,\quad  &\tcross\lessapprox t\lessapprox \tcorr, \\ \\
\frac{\sqrt{\pi}}{4}\,\left(\frac{\Delta_0}{2\,\nu}\right)^{3/2}\,\lambda\,b^{3/2}\,t\,,\quad  &\tcorr\ll t, \end{cases}
\end{split}\label{eq:VarScalingLeffLargerLeffC}
\end{align}
where, for $\tcorr\ll t$, we chose the exact numerical value of $\sqrt{\pi}/4$ for the universal amplitude $c_0$ \cite{KrugReview1997}. The behavior for $t\ll\tcross$ may be obtained in various ways. For one, we could take the short-time limit of \eqref{eq:ResultVarPsiOfU}. Alternatively, we know from the scaling arguments presented in Fig. \ref{fig:ScalingVarianceSchematic} that for these times the system is governed by the EW-scaling regime, which implies normal diffusive behavior according to the EW equation.\\
Hence, with the exact results in \autoref{sec:ExactResults} (see \eqref{eq:KPZTURIntermediate}) and the approximations for the variance we can formulate the TUR product $\mathcal{Q}$ in the long-time limit as
\begin{align}
\mathcal{Q}\simeq\left(5-\frac{1}{\Lambda}\right)\begin{cases} \left(1+\frac{\leff^2}{32\pi^2}\,\mathcal{S}_1(\Lambda)-\frac{\leff^4}{256\pi^4}\,\mathcal{S}_2(\Lambda)+O(\leff^6)\right)\,,\qquad&\text{for }\leff\lessapprox\leff^\text{c}, \\ \\
\frac{\sqrt{\pi}}{8\,\sqrt{2}}\,\leff\,,\qquad&\text{for }\leff^\text{c}\ll\leff. \end{cases}\label{eq:FullTUR}
\end{align}
Here we state the $\Lambda$-dependent result from \eqref{eq:VarOfPsiRes} in anticipation of the comparison to numerical simulations for a fixed system-size, which also implies a fixed value of $\Lambda$.

\section{Comparison with Numerical Simulations}\label{sec:CompNumSim}

\subsection{The Numerical Scheme}\label{subsec:NumComp_NumScheme}

In this section we present numerical simulations of \eqref{eq:KPZEq} via a stochastic Heun method as described in \cite{NiggemannSeifert2021}. We use these simulations to numerically determine the values of $\expval{\Psi(t)}^2$, $\stot$ and $\text{var}[\Psi(t)]$ and therefore $\mathcal{Q}$. Due to the sensitivity of the numerics to specific discretization of the KPZ non-linearity as discussed in \cite{NiggemannSeifert2021}, we here choose the discretization introduced in \cite{LamShin1998} (i.e., $\gamma=1/2$ in \cite{NiggemannSeifert2021}), as this proved to yield the most accurate results in \cite{NiggemannSeifert2021}. For the technical details and the respective definitions of the numerical observables we refer to \cite{NiggemannSeifert2021}. The numerical scheme uses scaled system parameters $\{\wt{\nu}\,,\wt{\Delta}_0\,,\wt{\lambda}\}$ given by
\begin{equation}
\wt{\nu}\equiv\frac{\nu}{\delta^2}\,,\qquad\wt{\Delta}_0\equiv\frac{\Delta_0}{\delta}\,,\qquad\wt{\lambda}\equiv\frac{\lambda}{\delta^2}\,\label{eq:DefNumSystemParameters}
\end{equation}
with $\delta=b/L$ and $L$ the numerical system-size \cite{NiggemannSeifert2021}. For the sake of simplicity, we set $\delta=1$. This implies for the effective coupling constant $\leff$,
\begin{equation}
\wt{\lambda}_\text{eff}=L^{1/2}\,\left(\frac{\wt{\Delta}_0}{\wt{\nu}^3}\right)^{1/2}\,\wt{\lambda}.\label{eq:DefNumLeff}
\end{equation}
For all numerical data shown here, we used $\wt{\nu}=\wt{\Delta}_0=1$ and thus the critical value of the effective coupling constant is reached for (see \eqref{eq:DefCriticalLeff})
\begin{equation}
\wt{\lambda}^\text{c}\approx\frac{12.28}{L^{1/2}}.\label{eq:DefNumCriticalLambda}
\end{equation}
In the case of $L=256$, which is the system-size we used for the data shown below, \eqref{eq:DefNumCriticalLambda} yields
\begin{equation}
\wt{\lambda}^\text{c}\approx0.768.\label{eq:DefNumCritLambdaL256}
\end{equation}
Like in \cite{NiggemannSeifert2021}, we use
\begin{equation}
\Lambda=\frac{L-1}{3},\label{eq:DefLambdaOfL}
\end{equation}
to establish a connection between the numerical system-size $L$ and the Fourier-cutoff parameter $\Lambda$ from e.g. \eqref{eq:FullTUR}. Thus, in terms of the numerical parameters $\{\wt{\nu}\,,\wt{\Delta}_0\,,\wt{\lambda}\}$ and $L$ the expression for $\expval{\Psi(t)}^2$ reads with \eqref{eq:ExactSSCurrent}
\begin{equation}
\expval{\Psi(t)}^2=\left(\frac{\wt{\Delta}_0\,\wt{\lambda}}{6\,\wt{\nu}}\,(L-1)\right)^2\,t^2,\label{eq:DefNumPsiSquared}
\end{equation}
and for $\stot$ we have with \eqref{eq:ExactSSSigma}
\begin{equation}
\stot=\frac{\wt{\Delta}_0}{36}\,\left(\frac{\wt{\lambda}}{\wt{\nu}}\right)^2\,\left(5\,L-13+\frac{8}{L}\right)\,t.\label{eq:DefNumStot}
\end{equation}
Both expressions in \eqref{eq:DefNumPsiSquared} and \eqref{eq:DefNumStot} may also be found in \cite{NiggemannSeifert2021}, however, there under the condition of $\leff\ll\leff^\text{c}$. Similarly, we get for system parameters, such that $\leff\lessapprox\leff^\text{c}$, and with the result from \eqref{eq:VarScalingLeffSmallerLeffC} the following expression for the variance of $\Psi(t)$,
\begin{equation}
\text{var}[\Psi(t)]\simeq\wt{\Delta}_0\,L\,t\,\left(1+L\,\frac{\wt{\Delta}_0\,\wt{\lambda}^2}{\wt{\nu}^3}\,\frac{\mathcal{S}_1(L)}{32\,\pi^2}-L^2\,\frac{\wt{\Delta}_0^2\,\wt{\lambda}^4}{\wt{\nu}^6}\,\frac{\mathcal{S}_2(L)}{256\,\pi^4}+O(\leff^6)\right),\label{eq:DefNumVarEWRegime}
\end{equation}
with
\begin{align}
\begin{split}
\mathcal{S}_1(L)&=2\sum_{k=1}^{\lceil \frac{L-1}{3} \rceil}\frac{1}{k^2}\,,\\
\mathcal{S}_2(L)&=2\sum_{k=1}^{\lceil \frac{L-1}{3} \rceil}\frac{1}{k^3}\sum_{m=-\lceil \frac{L-1}{3} \rceil+k}^{\lceil \frac{L-1}{3} \rceil}\frac{m}{k^2+(k-m)^2+m^2}-\sum_{k=1}^{\lceil \frac{L-1}{3} \rceil}\frac{1}{k^4}.
\end{split}\label{eq:DefSumsOfL}
\end{align}
On the other hand for parameter sets with $\leff^\text{c}\ll\leff$ we have with \eqref{eq:VarScalingLeffLargerLeffC}
\begin{align}
\begin{split}
\text{var}[\Psi(t)]\simeq\begin{cases}  
\wt{\Delta}_0\,L\,t\,,\quad  &t\ll\tcross, \\ \\
\frac{3\,\Gamma(2/3)}{8\,\pi^{2/3}}\,L\,\wt{\lambda}^{4/3}\,\left(\frac{\wt{\Delta}_0}{\wt{\nu}}\right)^{5/3}\,t^{4/3}\,,\quad  &\tcross\lessapprox t\lessapprox\tcorr, \\ \\
\frac{\sqrt{\pi}}{4}\,\left(\frac{\wt{\Delta}_0}{2\,\wt{\nu}}\right)^{3/2}\,\wt{\lambda}\,L^{3/2}\,t\,,\quad  &\tcorr\ll t\,. \end{cases}
\end{split}\label{eq:DefNumVarKPZRegime}
\end{align}
Accordingly, we have for the TUR product $\mathcal{Q}$
\begin{align}
\mathcal{Q}\simeq\left(5-\frac{3}{L-1}\right)\begin{cases} \left(1+L\,\frac{\wt{\Delta}_0\,\wt{\lambda}^2}{\wt{\nu}^3}\,\frac{\mathcal{S}_1(L)}{32\,\pi^2}-L^2\,\frac{\wt{\Delta}_0^2\,\wt{\lambda}^4}{\wt{\nu}^6}\,\frac{\mathcal{S}_2(L)}{256\,\pi^4}\right)\,,\qquad&\text{for }\leff\lessapprox\leff^\text{c}, \\ \\
\frac{\sqrt{\pi}}{8\,\sqrt{2}}\,\frac{\wt{\Delta}_0^{1/2}\,\wt{\lambda}}{\wt{\nu}^{3/2}}\,L^{1/2}\,,\qquad&\text{for }\leff^\text{c}\ll\leff. \end{cases}\label{eq:DefNumFullTUR}
\end{align}

\subsection{$\expval{\Psi(t)}^2$ and $\stot$}\label{subsec:NumComp_Psi2Stot}

\begin{figure}[tbph]
\centering
\begin{subfigure}{.45\textwidth}
\centering
\includegraphics[width=\textwidth]{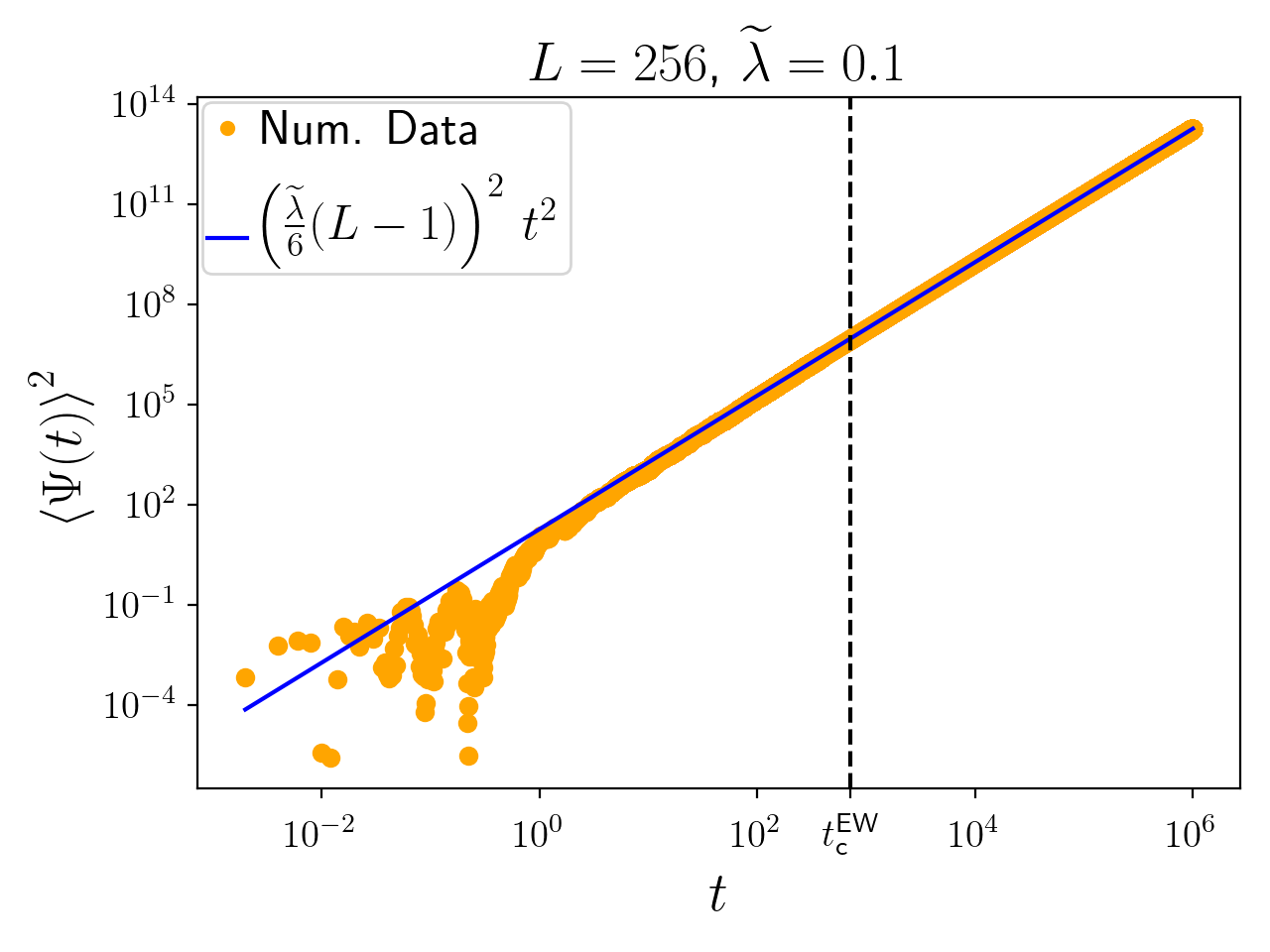}
\caption{}
\label{subfig:Psi2_lambda_1}
\end{subfigure}%
\hspace{0.1\textwidth}%
\begin{subfigure}{.45\textwidth}
\centering
\includegraphics[width=\textwidth]{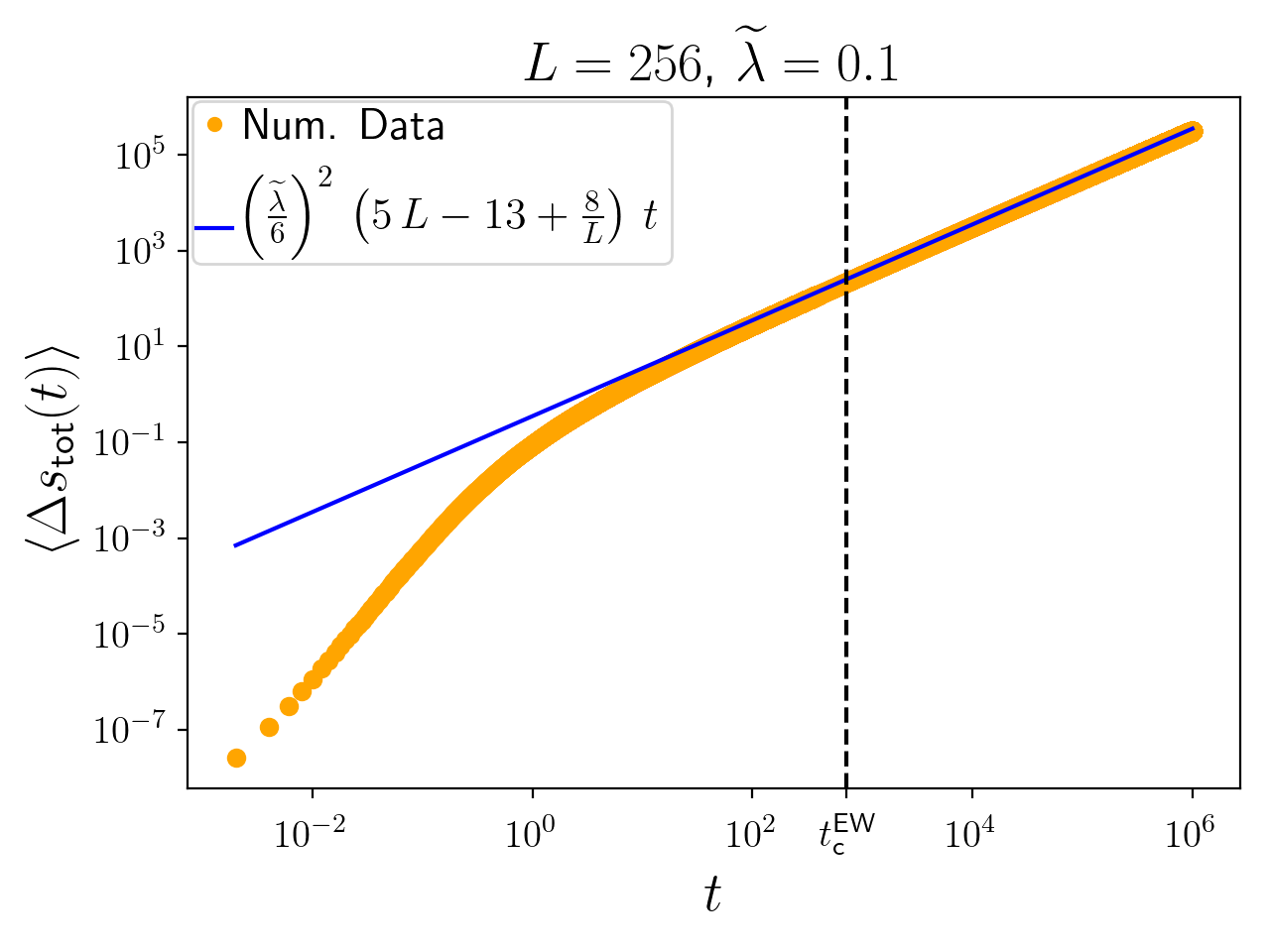}
\caption{}
\label{subfig:Stot_lambda_1}
\end{subfigure}
\begin{subfigure}{.45\textwidth}
\centering
\includegraphics[width=\textwidth]{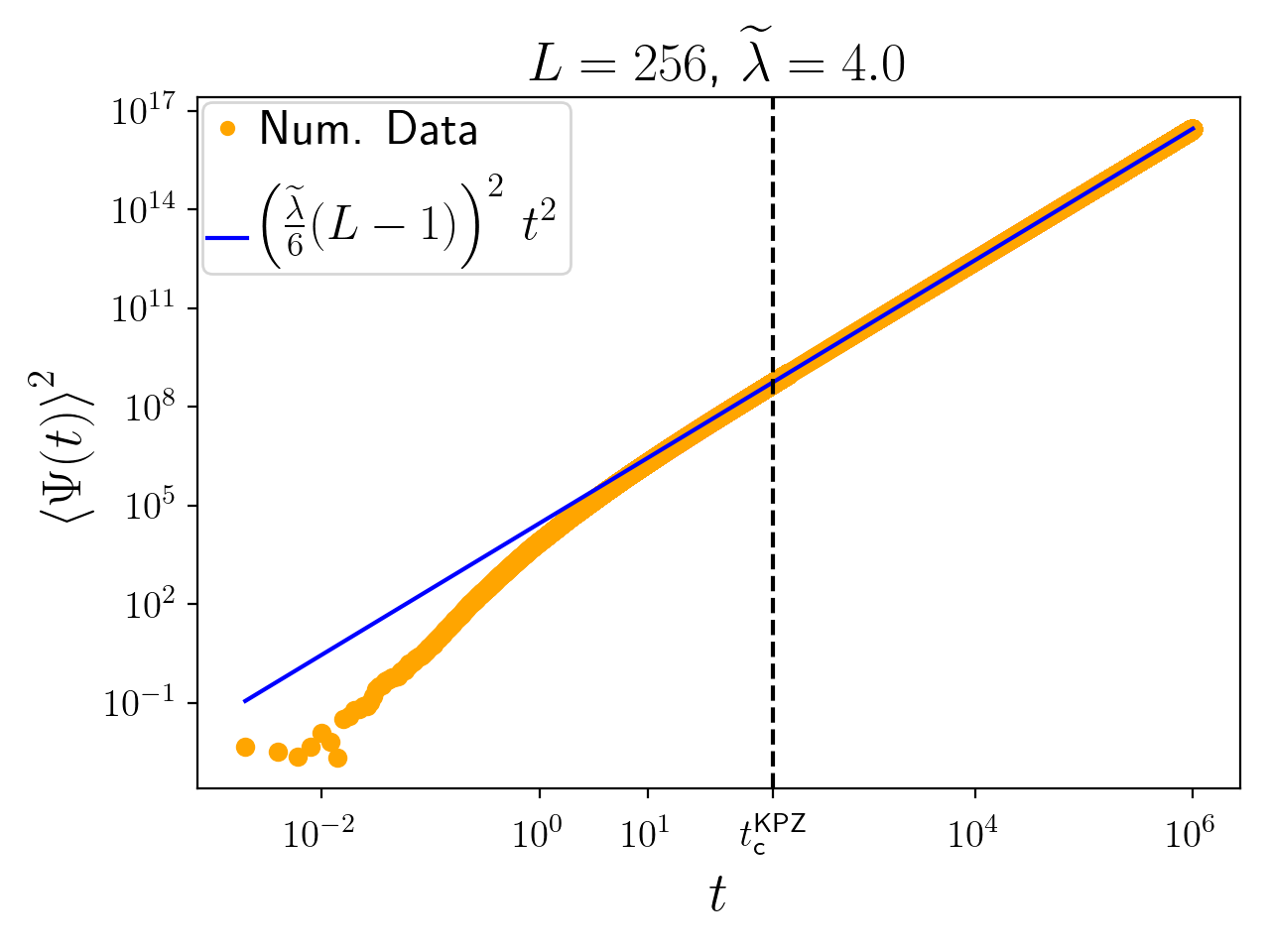}
\caption{}
\label{subfig:Psi2_lambda4}
\end{subfigure}%
\hspace{0.1\textwidth}%
\begin{subfigure}{.45\textwidth}
\centering
\includegraphics[width=\textwidth]{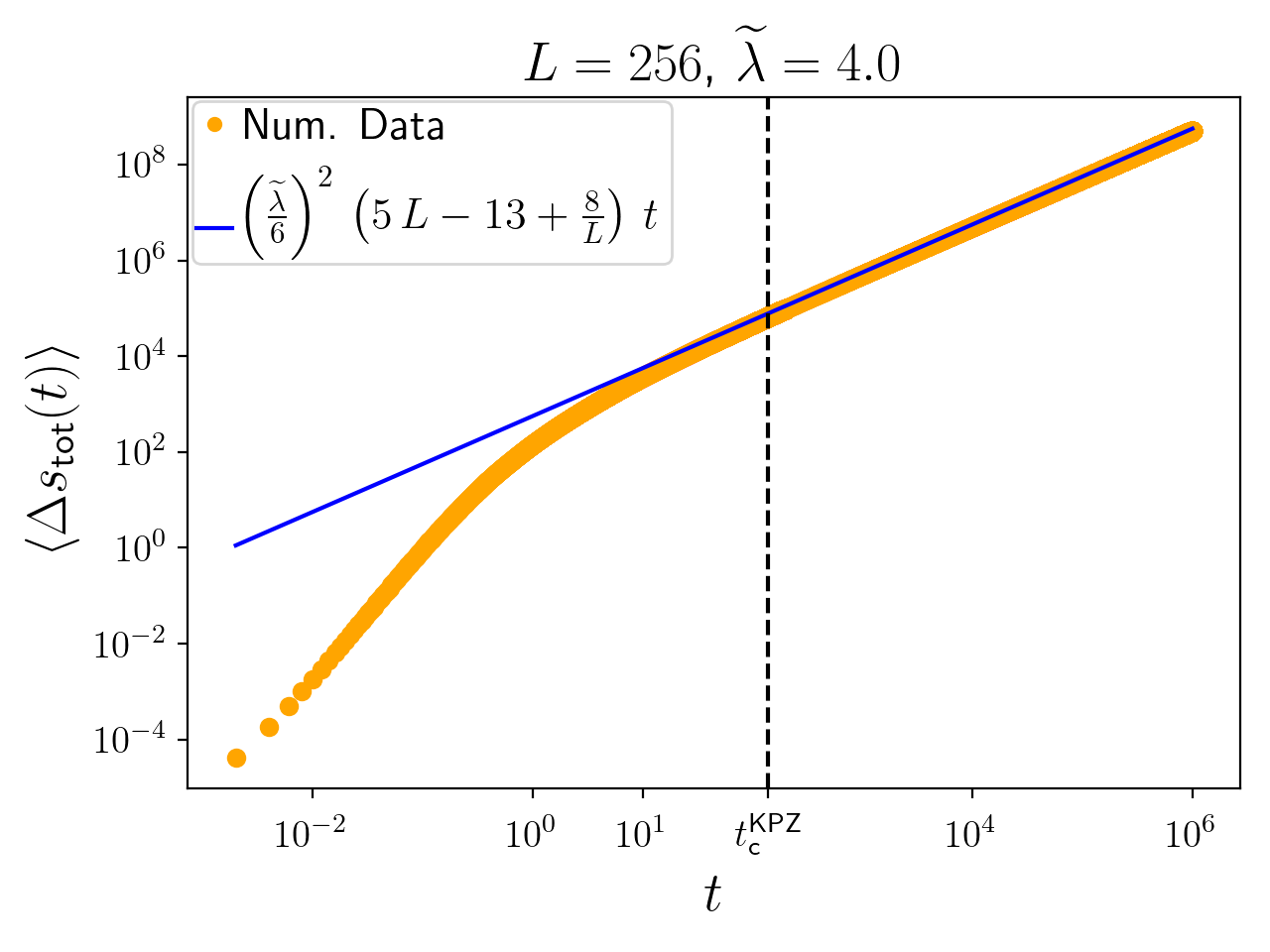}
\caption{}
\label{subfig:Stot_lambda4}
\end{subfigure}
\caption{Comparison of the numerical data obtained for a system-size of $L=256$ to $\expval{\Psi(t)}^2$ from \eqref{eq:DefNumPsiSquared} in (\subref{subfig:Psi2_lambda_1}) and (\subref{subfig:Psi2_lambda4}) and to $\stot$ from \eqref{eq:DefNumStot} in (\subref{subfig:Stot_lambda_1}) and (\subref{subfig:Stot_lambda4}), respectively. In (\subref{subfig:Psi2_lambda_1}) and (\subref{subfig:Stot_lambda_1}) we simulate a system with $\wt{\lambda}=0.1<\wt{\lambda}^\text{c}$, i.e., in the EW scaling regime of the KPZ equation, whereas in (\subref{subfig:Psi2_lambda4}) and (\subref{subfig:Stot_lambda4}) we simulate a system with $\wt{\lambda}=4>\wt{\lambda}^\text{c}$, which puts the system in the KPZ scaling regime. The vertical lines indicate the respective correlation times $\tcEW$ from \eqref{eq:EWCorrTime} and $\tcorr$ from \eqref{eq:KPZCorrelationTime}.}
\label{fig:Psi2Stot}
\end{figure}
In Fig. \ref{fig:Psi2Stot} we show for two values of $\wt{\lambda}$ a comparison of numerical data for both $\expval{\Psi(t)}^2$ and $\stot$ with the respective theoretical predictions according to \eqref{eq:DefNumPsiSquared} and \eqref{eq:DefNumStot}. In the case of $\wt{\lambda}=0.1$ the system is in the EW scaling regime of the KPZ equation and thus the relevant time-scale is the EW correlation time $\tcEW$, which is indicated by the vertical line in Figs. \ref{fig:Psi2Stot}(\subref{subfig:Psi2_lambda_1}) and (\subref{subfig:Stot_lambda_1}). As can be seen well, for times $t>\tcEW$ the numerical data follows the theoretical prediction for both $\expval{\Psi(t)}^2$ and $\stot$. For $\wt{\lambda}=4.0$ the system is in its KPZ scaling regime, which implies that the numerical data is expected to converge to the theoretical predictions for times $t>\tcorr$, i.e., the KPZ correlation time. In Figs. \ref{fig:Psi2Stot}(\subref{subfig:Psi2_lambda4}) and (\subref{subfig:Stot_lambda4}) this convergence can be well observed. Thus, the results in Fig. \ref{fig:Psi2Stot} are additional support for the fact that the expressions for $\expval{\Psi(t)}$ and $\stot$ obtained analytically in \eqref{eq:DefNumPsiSquared} and \eqref{eq:DefNumStot}, respectively, hold for an arbitrary coupling constant. This extends the range of validity for these two entities from the EW regime (or the weak-coupling limit) (see \cite{NiggemannSeifert2020,NiggemannSeifert2021}) to the KPZ regime (or the strong-coupling limit).

\subsection{Variance of $\Psi(t)$ and Universal Scaling Amplitude}\label{subsec:NumComp_Var}

\begin{figure}[tbph]
\centering
\begin{subfigure}{.45\textwidth}
\centering
\includegraphics[width=\textwidth]{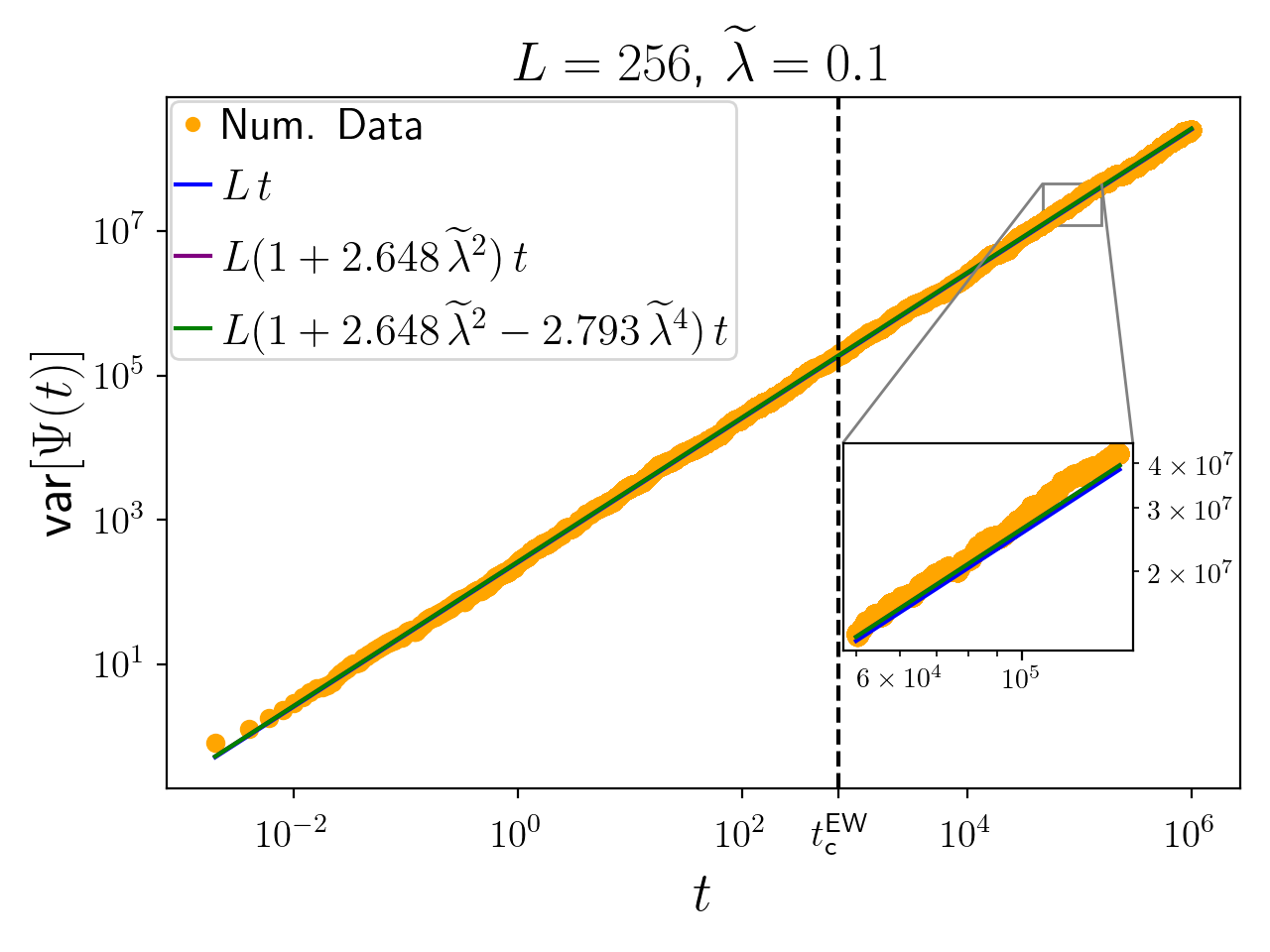}
\caption{}
\label{subfig:Var_lambda_1}
\end{subfigure}%
\hspace{0.1\textwidth}%
\begin{subfigure}{.45\textwidth}
\centering
\includegraphics[width=\textwidth]{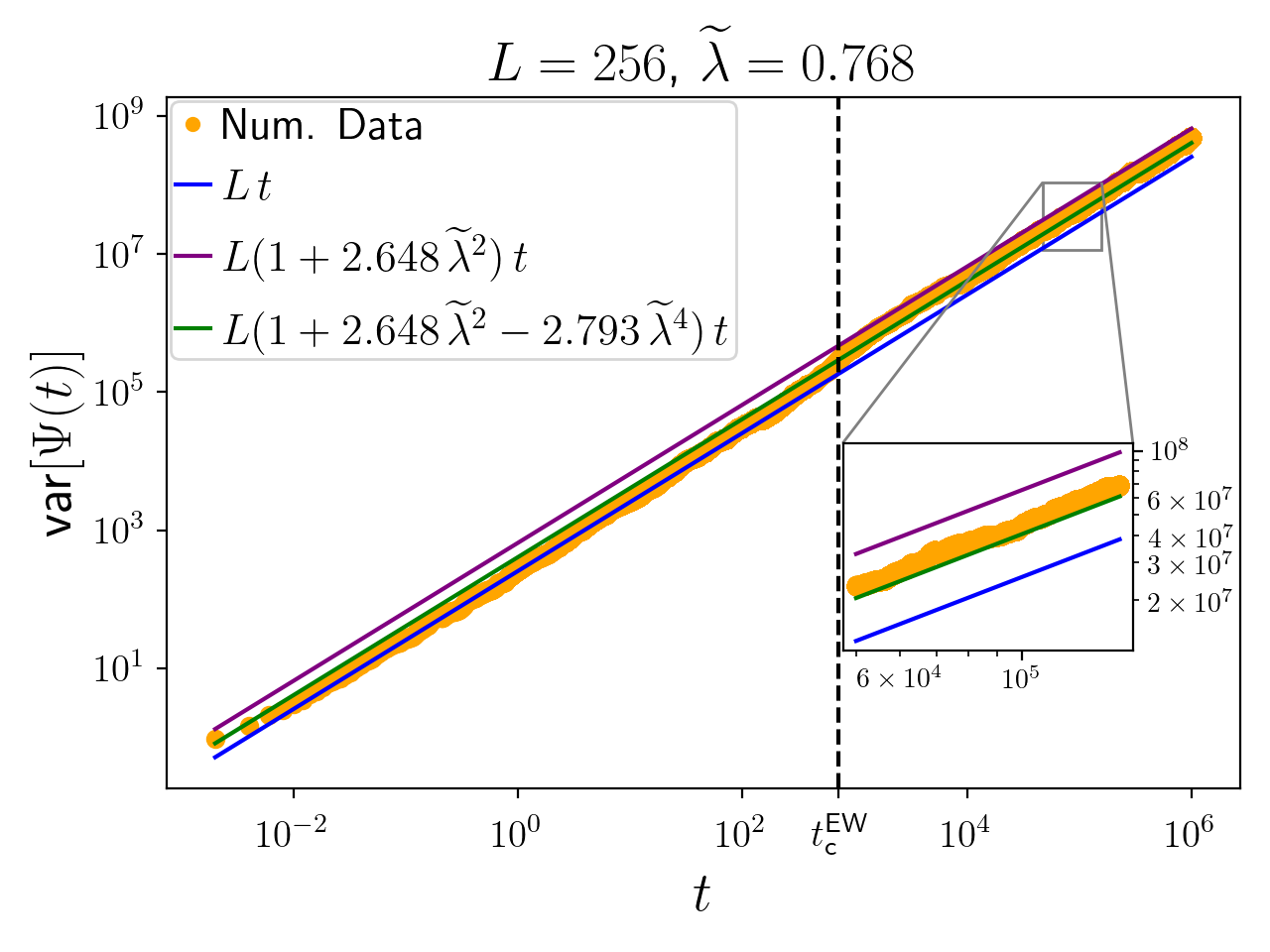}
\caption{}
\label{subfig:Var_lambda_768}
\end{subfigure}
\caption{Comparison of numerical data obtained for $L=256$ with $\wt{\lambda}=0.1$ in (\subref{subfig:Var_lambda_1}) and $\wt{\lambda}=0.768\approx\wt{\lambda}^\text{c}$ in (\subref{subfig:Var_lambda_768}) with the theoretical prediction from \eqref{eq:DefNumVarEWRegime} of the variance of $\Psi(t)$ in the EW scaling regime. We show three orders of approximation, i.e., $\wt{\lambda}^0$, $\wt{\lambda}^2$ and $\wt{\lambda}^4$, for $\text{var}[\Psi(t)]$ to demonstrate the effect of including higher order terms. The vertical line indicates the EW correlation time $\tcEW$ from \eqref{eq:EWCorrTime}.}
\label{fig:VarEWRegime}
\end{figure}
Fig. \ref{fig:VarEWRegime} shows numerically obtained data of the variance of $\Psi(t)$ for a system-size of $L=256$ and for $\wt{\lambda}=0.1$ (see Fig. \ref{fig:VarEWRegime}(\subref{subfig:Var_lambda_1})), $\wt{\lambda}=0.768\approx\wt{\lambda}^\text{c}$ from \eqref{eq:DefNumCritLambdaL256} (see Fig. \ref{fig:VarEWRegime}(\subref{subfig:Var_lambda_768})). To demonstrate the effect of including higher order terms in the approximation of $\text{var}[\Psi(t)]$ in the EW scaling regime, we show in Fig. \ref{fig:VarEWRegime} each partial sum of the expansion in \eqref{eq:DefNumVarEWRegime} separately in increasing order. As can be seen clearly in Fig. \ref{fig:VarEWRegime}(\subref{subfig:Var_lambda_1}), there is no discernible difference between the lowest and highest order perturbation result for $\wt{\lambda}=0.1$. In Fig. \ref{fig:VarEWRegime}(\subref{subfig:Var_lambda_768}), for $\wt{\lambda}=0.768$, however, the difference between the three approximation orders becomes apparent. Here the zero-order approximation ($\wt{\lambda}^0$) underestimates the numerical data and the first-order approximation ($\wt{\lambda}^2$) is a slight overestimation, whereas the second-order result ($\wt{\lambda}^4$) matches the numerical data well. For values $\wt{\lambda}>\wt{\lambda}^\text{c}$, we leave the region in which the perturbation expansion from \autoref{sec:PerturbationExpansion} is expected to be valid, which is reflected in a rapid decline in the quality of the highest-order approximation (not shown explicitly), as is to be expected. The numerical values in the legend of Fig. \ref{fig:VarEWRegime} are obtained by evaluating \eqref{eq:DefSumsOfL} and inserting these results into \eqref{eq:DefNumVarEWRegime} for $L=256$ (i.e., $\lceil (L-1)/3\rceil=85$).
\begin{figure}[tbhp]
\centering
\includegraphics[width=0.66\textwidth]{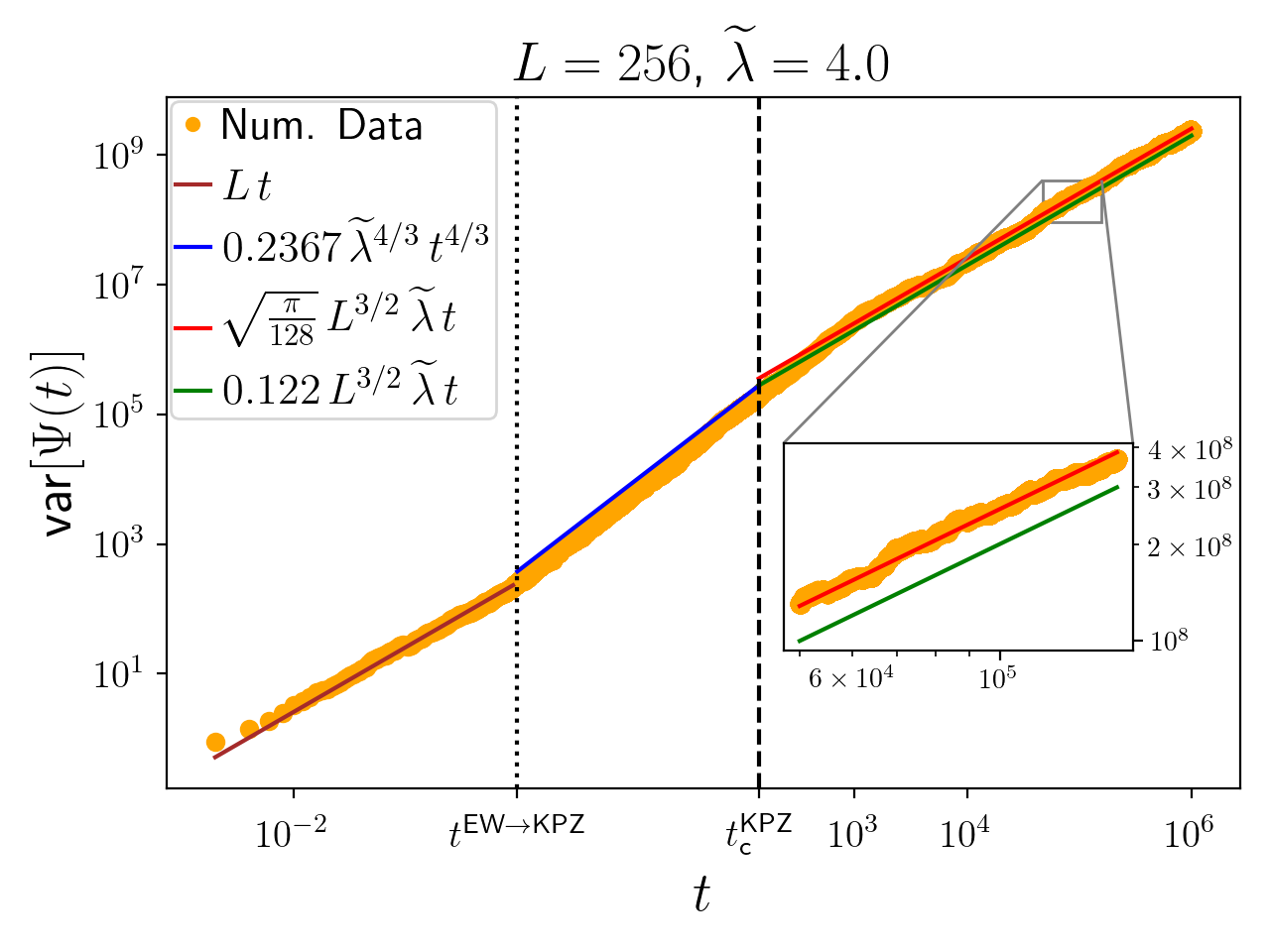}
\caption{Numerical data of $\text{var}[\Psi(t)]$ for $\wt{\lambda}=4.0>\wt{\lambda}^\text{c}$, i.e., in the KPZ regime. For $t<\tcross$ we plot the variance according to the first line in \eqref{eq:DefNumVarKPZRegime}. In the transient regime, i.e., for $\tcross<t<\tcorr$, the variance according to the second line in \eqref{eq:DefNumVarKPZRegime} is shown. For $t>\tcorr$, i.e., in the stationary KPZ regime, we plot for one the third line from \eqref{eq:DefNumVarKPZRegime} and on the other hand the second line from \eqref{eq:ResultMatchingConstant1}.}\label{fig:VarKPZRegime}
\end{figure}
For the case of $\leff^\text{c}<\leff$, we show in Fig. \ref{fig:VarKPZRegime} numerical data of the variance of $\Psi(t)$. As can be seen clearly, the variance displays the expected scaling behaviors (see \eqref{eq:DefNumVarKPZRegime}), namely, on the one hand, for times $t<\tcross$ scaling according to the EW scaling regime of the KPZ equation. On the other hand, for times $t>\tcross$ Fig. \ref{fig:VarKPZRegime} shows the typical KPZ scaling regime behavior, namely for $\tcross<t<\tcorr$ the transient regime with its super-diffusivity and for $\tcorr<t$ the stationary KPZ regime. In regard of the EW to KPZ crossover time from \eqref{eq:EWtoKPZCrossoverTime}, we see very good agreement between the theoretical prediction, indicated by the left vertical line in Fig. \ref{fig:VarKPZRegime} and the numerical data. However, the theoretical prediction for the KPZ correlation time from \eqref{eq:KPZCorrelationTime}, shown as the right vertical line in Fig. \ref{fig:VarKPZRegime} seems to be too small, as the super-diffusive behavior continues beyond $\tcorr$. We will investigate this in more detail below. This discrepancy aside, we find good agreement in all three sub-regimes of the variance between the numerical data and the theoretical predictions from \eqref{eq:ResultMatchingConstant1} and \eqref{eq:DefNumVarKPZRegime}.
\begin{figure}[tbhp]
\centering
\includegraphics[width=0.66\textwidth]{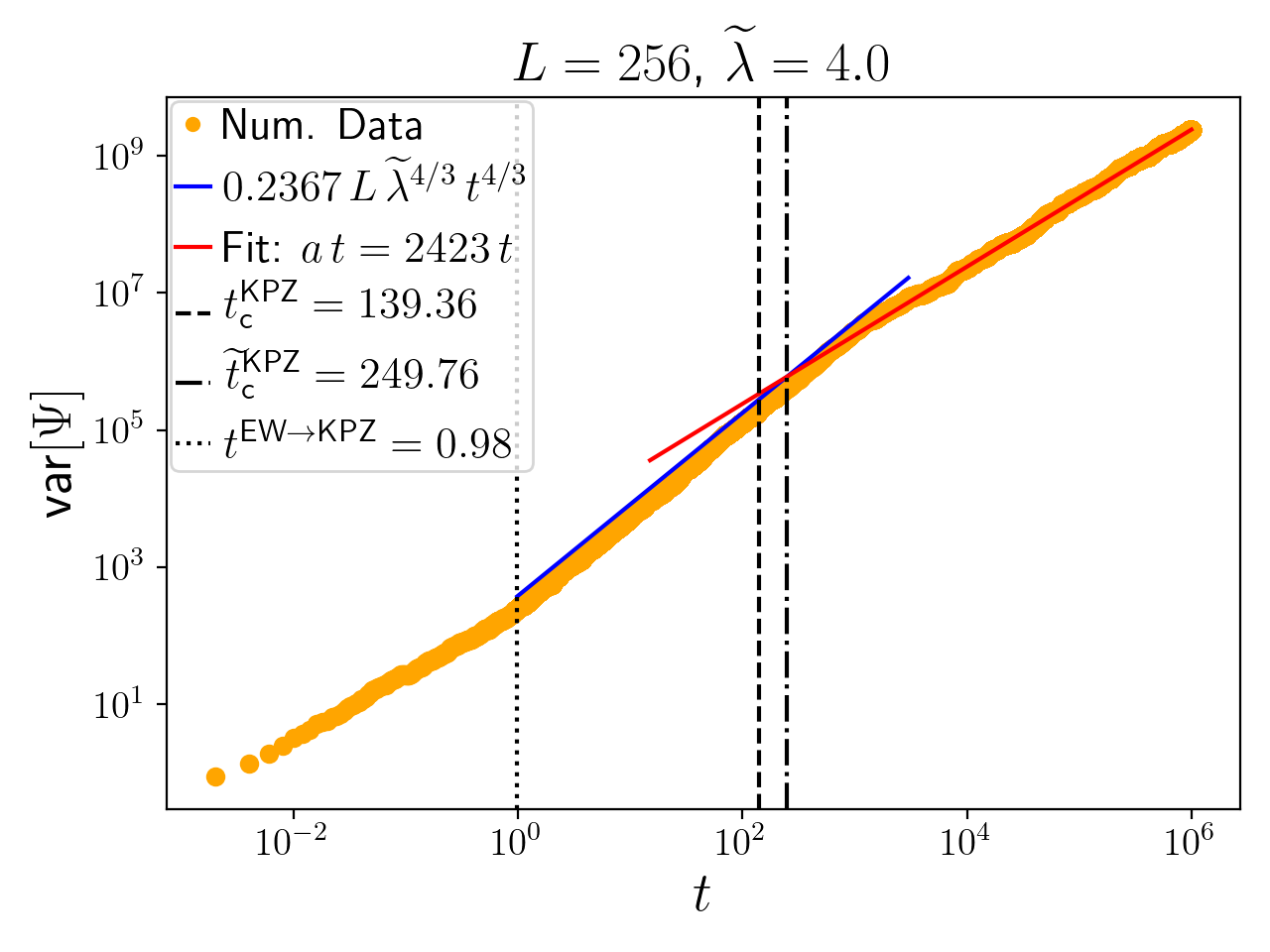}
\caption{Exemplary determination of the numerical KPZ correlation time $\wt{t}_\text{c}^\text{KPZ}$ for $\wt{\lambda}=4.0$.}
\label{fig:TcEstimation}
\end{figure}
In Fig. \ref{fig:TcEstimation} we show our approach to determining the numerical KPZ correlation time $\wt{t}_\text{c}^\text{KPZ}$. In particular, we search the time at which the transient behavior according to \eqref{eq:DefNumVarKPZRegime} becomes equal to the stationary branch. Here we determine the latter by fitting the numerical data in the stationary KPZ regime via a fit-function $a\,t$, with $a$ the fit-parameter. Hence, we find for $\wt{t}_\text{c}^\text{KPZ}$,
\begin{equation}
\wt{t}_\text{c}^\text{KPZ}=\left(\frac{8\,\pi^{2/3}\,\wt{\nu}^{5/3}\,a}{3\,\Gamma(2/3)\,\wt{\Delta}_0^{5/3}\,\wt{\lambda}^{4/3}\,L}\right)^3.\label{eq:DefNumTcKPZ}
\end{equation}
Tab. \ref{tab:TcOfLambda} shows the numerically obtained $\wt{t}_\text{c}^\text{KPZ}$ and $\tcorr$ from \eqref{eq:KPZCorrelationTime} in dependence of $\wt{\lambda}$. For all values of $\wt{\lambda}$ the numerically obtained correlation time $\wt{t}_\text{c}^\text{KPZ}$ is roughly a factor of $2$ larger than the one from \eqref{eq:KPZCorrelationTime}. To be precise
\begin{equation}
\wt{t}_\text{c}^\text{KPZ}=(2.18\pm0.46)\,\tcorr,\label{eq:NumTcKPZResult}
\end{equation}
where the factor is the mean of the right hand column of Tab. \ref{tab:TcOfLambda} and the error the standard deviation of the mean. Let us use \eqref{eq:NumTcKPZResult} to reevaluate the universal scaling amplitude $c_0$ from \eqref{eq:KrugVar} according to the calculation in \eqref{eq:ResultMatchingConstant1}, which leads to
\begin{align}
\begin{split}
\mathcal{A}\,\wt{t}_\text{c}^\text{KPZ}&=\frac{3\,\Gamma(2/3)}{8\,\pi^{2/3}}\,\frac{\wt{\Delta_0}^{5/3}\,\wt{\lambda}^{4/3}}{\wt{\nu}^{5/3}}\,L\,\left(\wt{t}_\text{c}^\text{KPZ}\right)^{4/3}\,,\\
\mathcal{A}&=(0.45\pm0.03)\,\left(\frac{\wt{\Delta}_0}{2\,\wt{\nu}}\right)^{3/2}\,\wt{\lambda}\,L^{1/2}\,.
\end{split}\label{eq:NumC0AmplitudeCalculation}
\end{align}
Hence, we get for the universal scaling amplitude $c_0$,
\begin{equation}
c_0=(0.45\pm0.03),\label{eq:NumC0AmplitudeResult}
\end{equation}
where the theoretically predicted value from \cite{KrugReview1997} is $\sqrt{\pi}/4\approx0.44$, which is well inside the error bars of \eqref{eq:NumC0AmplitudeResult}. Thus, by using the numerically obtained value of the KPZ correlation time, $\wt{t}_\text{c}^\text{KPZ}$ from \eqref{eq:NumTcKPZResult}, and the DRG result for the variance of $\Psi(t)$ from \eqref{eq:DefNumVarKPZRegime} in the transient regime with the matching condition from \eqref{eq:MatchingConditionAtCorrelationTime} we are able to obtain the universal scaling amplitude from \eqref{eq:KrugVar} to good accuracy (see \eqref{eq:NumC0AmplitudeResult}). The result in \eqref{eq:NumC0AmplitudeResult} is a considerable improvement of \eqref{eq:ResultMatchingConstant1} which used $\tcorr$ from \eqref{eq:KPZCorrelationTime}.
\begin{table}[tbhp]
\centering
\caption{Numerical Estimation of $\wt{t}_\text{c}^\text{KPZ}$}
\begin{tabular}{c c c c}
$\wt{\lambda}$ & $\tcorr$ & $\wt{t}_\text{c}^\text{KPZ}$ & $\wt{t}_\text{c}^\text{KPZ}/\tcorr$ \\
$0.87$ & $640.74$ & $1839.26$ & $2.87$ \\
$1.0$ & $557.45$ & $1446.18$ & $2.59$ \\
$1.25$ & $445.96$ & $745.42$ & $1.67$ \\
$1.5$ & $371.63$ & $1032.80$ & $2.78$ \\
$1.75$ & $318.54$ & $584.15$ & $1.83$ \\
$2.0$ & $278.72$ & $524.67$ & $1.88$ \\
$2.5$ & $222.98$ & $496.44$ & $2.23$ \\
$3.0$ & $185.82$ & $363.50$ & $1.96$ \\
$4.0$ & $139.36$ & $249.76$ & $1.79$ \\\\
\end{tabular}
\label{tab:TcOfLambda}
\caption*{Numerical values of the theoretically predicted KPZ correlation time $\tcorr$ from \eqref{eq:KPZCorrelationTime} and the numerically obtained $\wt{t}_\text{c}^\text{KPZ}$ (see Fig. \ref{fig:TcEstimation}) in dependence of $\wt{\lambda}$ and for $L=256$. The right column shows the ratio of the two correlation times.}
\end{table}

\subsection{TUR Product $\mathcal{Q}$}\label{subsec:NumComp_Q}

\begin{figure}[tbph]
\centering
\begin{subfigure}{.45\textwidth}
\centering
\includegraphics[width=\textwidth]{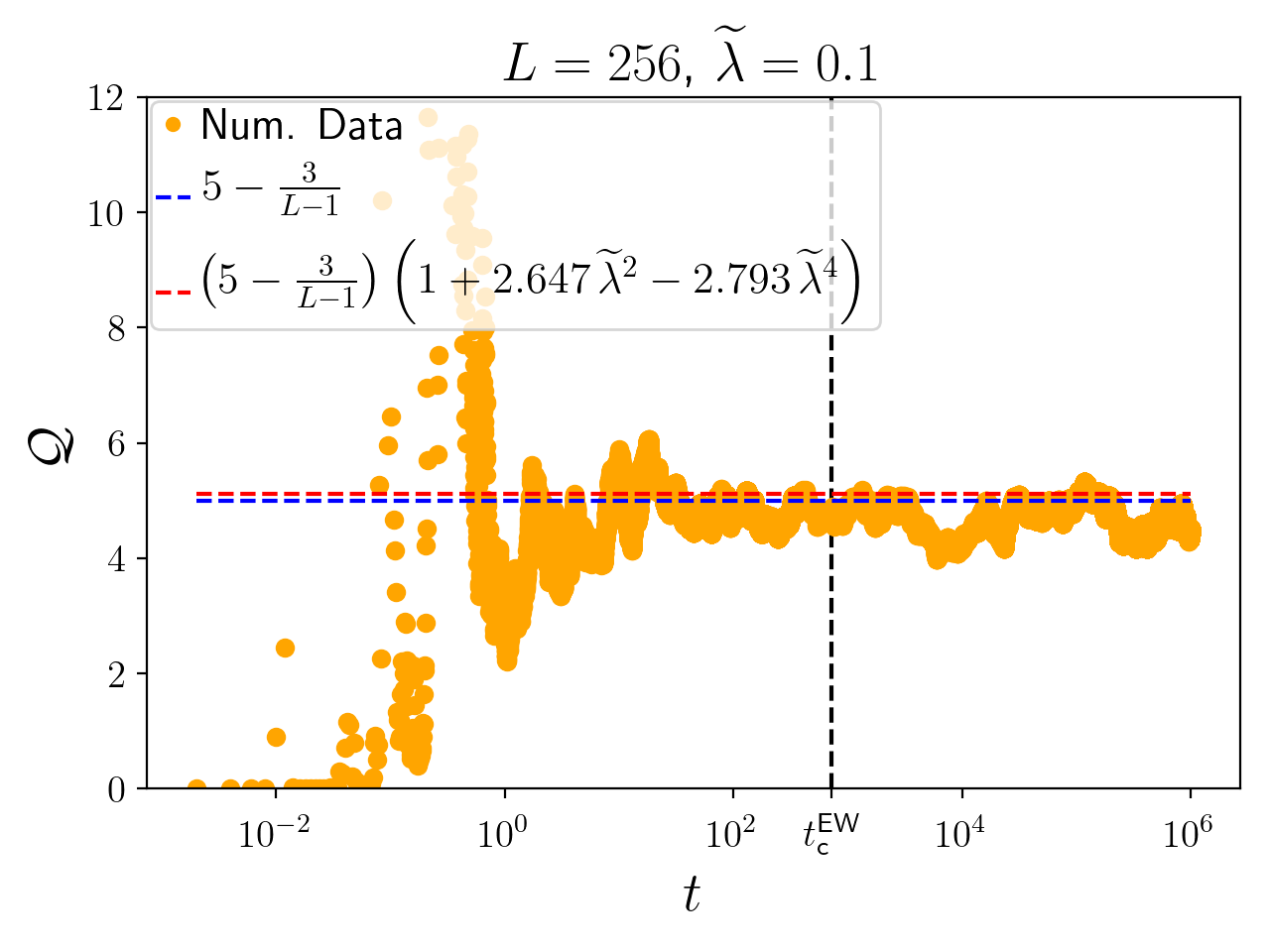}
\caption{}
\label{subfig:TUR_lambda_1}
\end{subfigure}%
\hspace{0.1\textwidth}%
\begin{subfigure}{.45\textwidth}
\centering
\includegraphics[width=\textwidth]{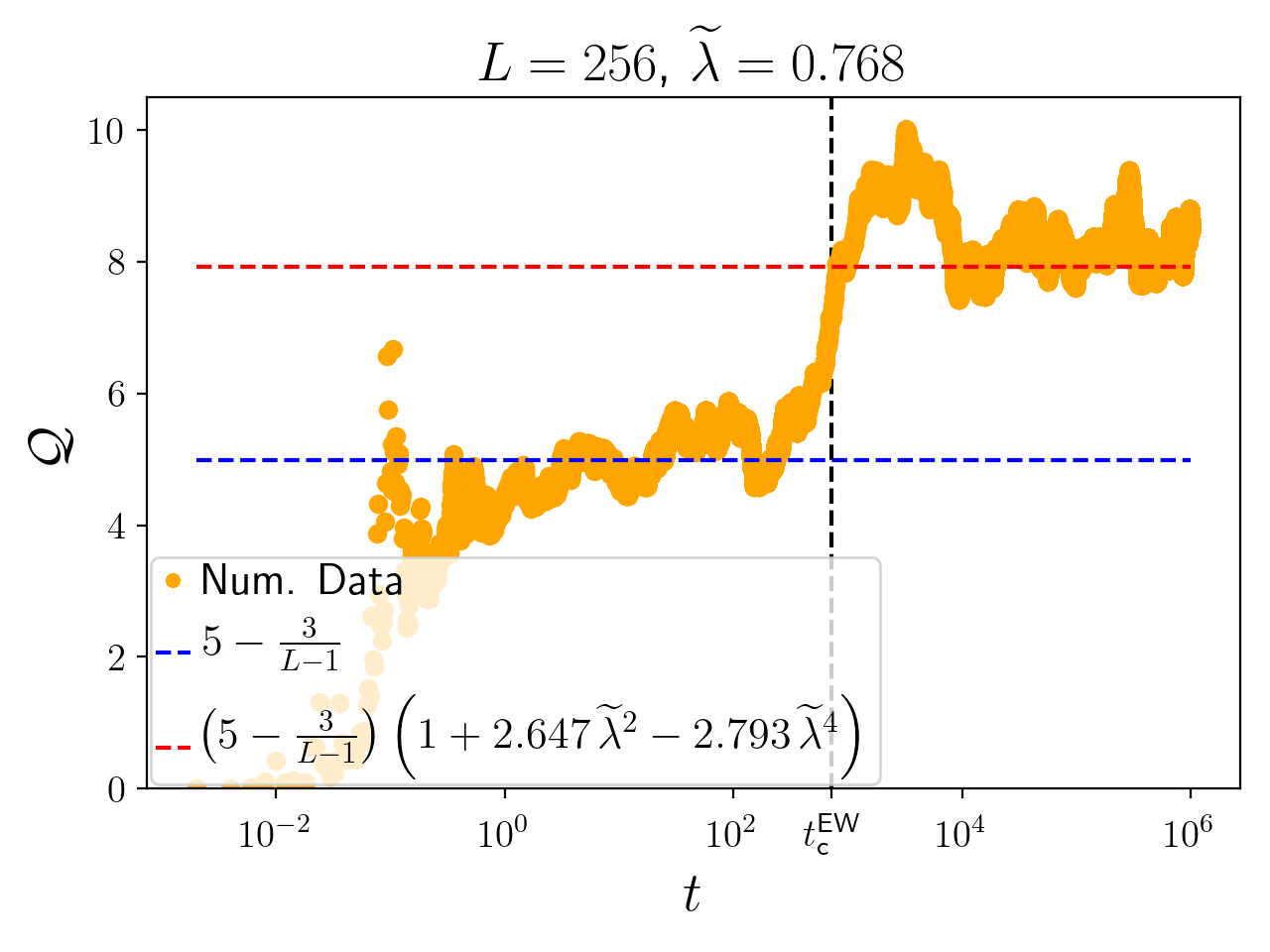}
\caption{}
\label{subfig:TUR_lambda_768}
\end{subfigure}
\begin{subfigure}{0.45\textwidth}
\centering
\includegraphics[width=\textwidth]{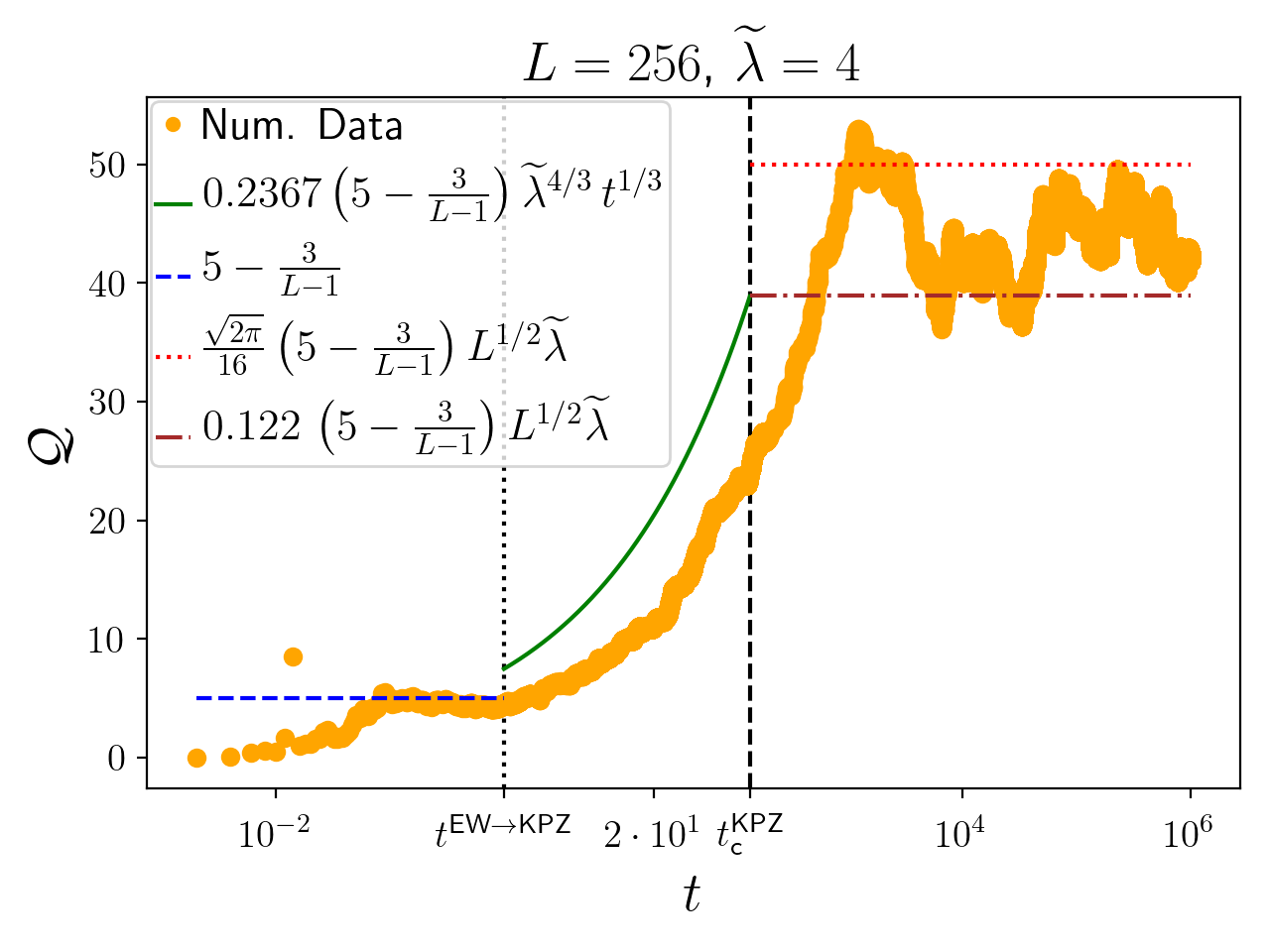}
\caption{}
\label{subfig:TUR_lambda4}
\end{subfigure}%
\hspace{0.1\textwidth}%
\begin{subfigure}{0.45\textwidth}
\centering
\includegraphics[width=\textwidth]{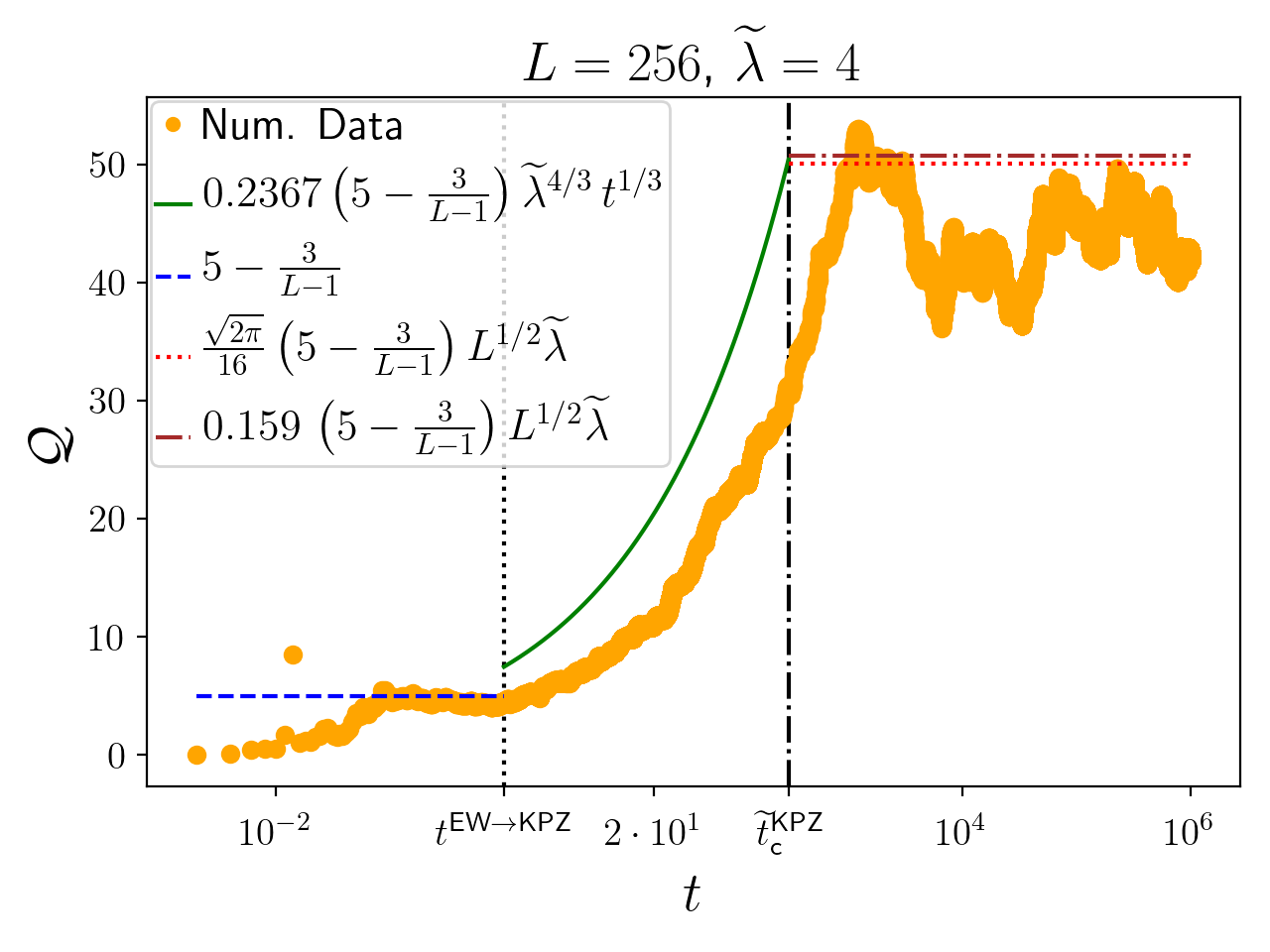}
\caption{}
\label{subfig:TUR_lambda4_corr}
\end{subfigure}%
\caption{Comparison of numerical data obtained for $L=256$ and $\wt{\lambda}=0.1$, $\wt{\lambda}=0.768\approx\wt{\lambda}^\text{c}$ and $\wt{\lambda}=4.0$ with the theoretical prediction of the TUR product $\mathcal{Q}$ both in the EW scaling regime of the KPZ equation, (\subref{subfig:TUR_lambda_1}) and (\subref{subfig:TUR_lambda_768}), as well as in the KPZ regime, (\subref{subfig:TUR_lambda4}), (\subref{subfig:TUR_lambda4_corr}). Note, that in (\subref{subfig:TUR_lambda4_corr}) we use the numerically obtained values for $\wt{t}^\text{KPZ}_\text{c}$ from \eqref{eq:NumTcKPZResult} and for $c_0$ from \eqref{eq:NumC0AmplitudeResult}.}
\label{fig:TUROfT}
\end{figure}
In Fig. \ref{fig:TUROfT} we show for three specific values of $\wt{\lambda}$ the time-evolution of the TUR product $\mathcal{Q}$. As can be seen for the two cases of $\wt{\lambda}\leq\wt{\lambda}^\text{c}$, Fig. \ref{fig:TUROfT}(\subref{subfig:TUR_lambda_1}), (\subref{subfig:TUR_lambda_768}), the perturbation expansion from \eqref{eq:DefNumFullTUR} yields convincing agreement with the numerical data for times $t\geq\tcEW$. To demonstrate the effect of the higher order contributions in the perturbation scheme, we also plot the zero-order result for reference. In Fig. \ref{fig:TUROfT}(\subref{subfig:TUR_lambda4}), (\subref{subfig:TUR_lambda4_corr}), i.e., in the KPZ scaling regime, we find that for times $t\leq\tcross$ the TUR product $\mathcal{Q}$ converges to the EW scaling result, namely $\mathcal{Q}=5-3/(L-1)$. For times $t\geq\tcorr$ we see the final convergence to the KPZ steady state result of $\mathcal{Q}$, where in Fig. \ref{fig:TUROfT}(\subref{subfig:TUR_lambda4}) the upper horizontal line indicates $\mathcal{Q}$ according to \eqref{eq:DefNumFullTUR} and the lower line represents the result one obtains by using \eqref{eq:ResultMatchingConstant1}. Both can be seen as reasonable approximations to the steady state value of $\mathcal{Q}$, however, in the light of the results regarding the numerical correlation time $\wt{t}^\text{KPZ}_\text{c}$ and the then resulting universal scaling amplitude in \eqref{eq:NumC0AmplitudeResult}, we regard the upper line as the more reliable one. This is further supported by the observation in \cite{NiggemannSeifert2021} that the numerical scheme intrinsically underestimates the TUR product $\mathcal{Q}$. For the theoretical prediction in the transient regime of the KPZ equation in Fig. \ref{fig:TUROfT}(\subref{subfig:TUR_lambda4}) we rely on the assumption that the steady-state results of $\expval{\Psi(t)}^2$ from \eqref{eq:DefNumPsiSquared} and $\stot$ from \eqref{eq:DefNumStot} yield reasonable approximations even for times $t$ smaller than the KPZ correlation time $\tcorr$. This is to some extent justified by the findings in Fig. \ref{fig:Psi2Stot}. We thus expect for $\tcross<t<\tcorr$ using \eqref{eq:DefNumVarKPZRegime},
\begin{equation}
\mathcal{Q}=\left(5-\frac{3}{L-1}\right)\,\frac{3\,\Gamma(2/3)}{8\,\pi^{2/3}}\,\frac{\wt{\Delta}_0^{2/3}}{\wt{\nu}^{5/3}}\,\wt{\lambda}^{4/3}\,t^{1/3},\label{eq:DefNumQTransient}
\end{equation}
which is what we plotted in Fig. \ref{fig:TUROfT}(\subref{subfig:TUR_lambda4}). As can be seen, the expression in \eqref{eq:DefNumQTransient} predicts the transient time-behavior well. The slight offset may either be a result of the intrinsic numerical underestimation of $\mathcal{Q}$ \cite{NiggemannSeifert2021} or originate in a minor error in the DRG result from \eqref{eq:DefNumVarKPZRegime} in terms of the numerical prefactor $c_0$. In Fig. \ref{fig:TUROfT}(\subref{subfig:TUR_lambda4_corr}), we show the same graphs as in Fig. \ref{fig:TUROfT}(\subref{subfig:TUR_lambda4}). However, here we use the numerically obtained value of the KPZ correlation time, $\wt{t}^\text{KPZ}_\text{c}$ from \eqref{eq:NumTcKPZResult}, and the corresponding reevaluated universal scaling amplitude from \eqref{eq:NumC0AmplitudeResult}, which replaces the numerical prefactor in \eqref{eq:ResultMatchingConstant1}. This leads to the closing of the gap between the two stationary results in the KPZ regime and thus smoothes the transition between the two branches of \eqref{eq:DefNumVarKPZRegime} for times $t>\tcross$.
\begin{figure}[tbhp]
\centering
\begin{subfigure}{.45\textwidth}
\centering
\includegraphics[width=\textwidth]{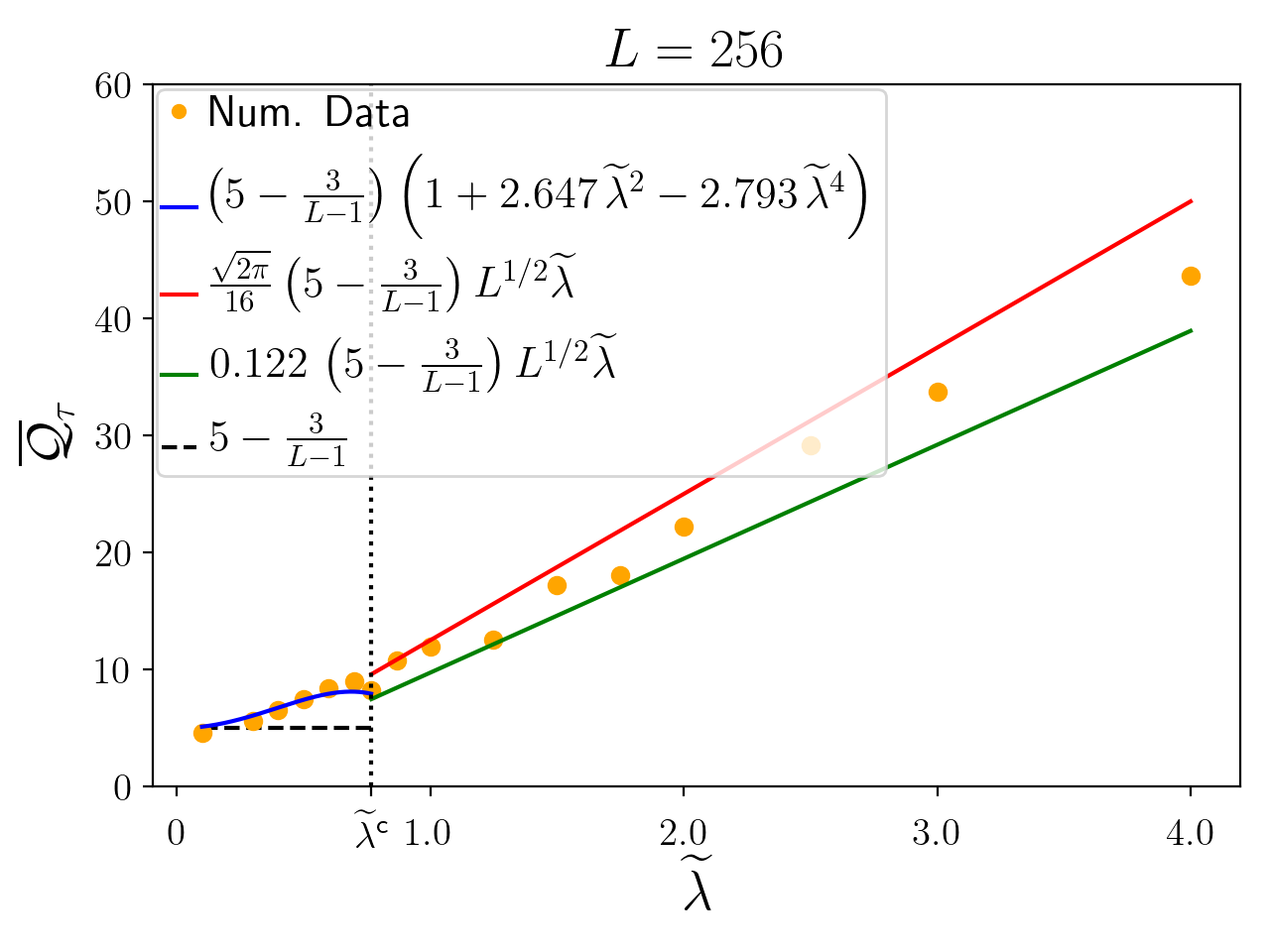}
\caption{}
\label{subfig:TUROfLambda}
\end{subfigure}%
\hspace{0.1\textwidth}%
\begin{subfigure}{.45\textwidth}
\centering
\includegraphics[width=\textwidth]{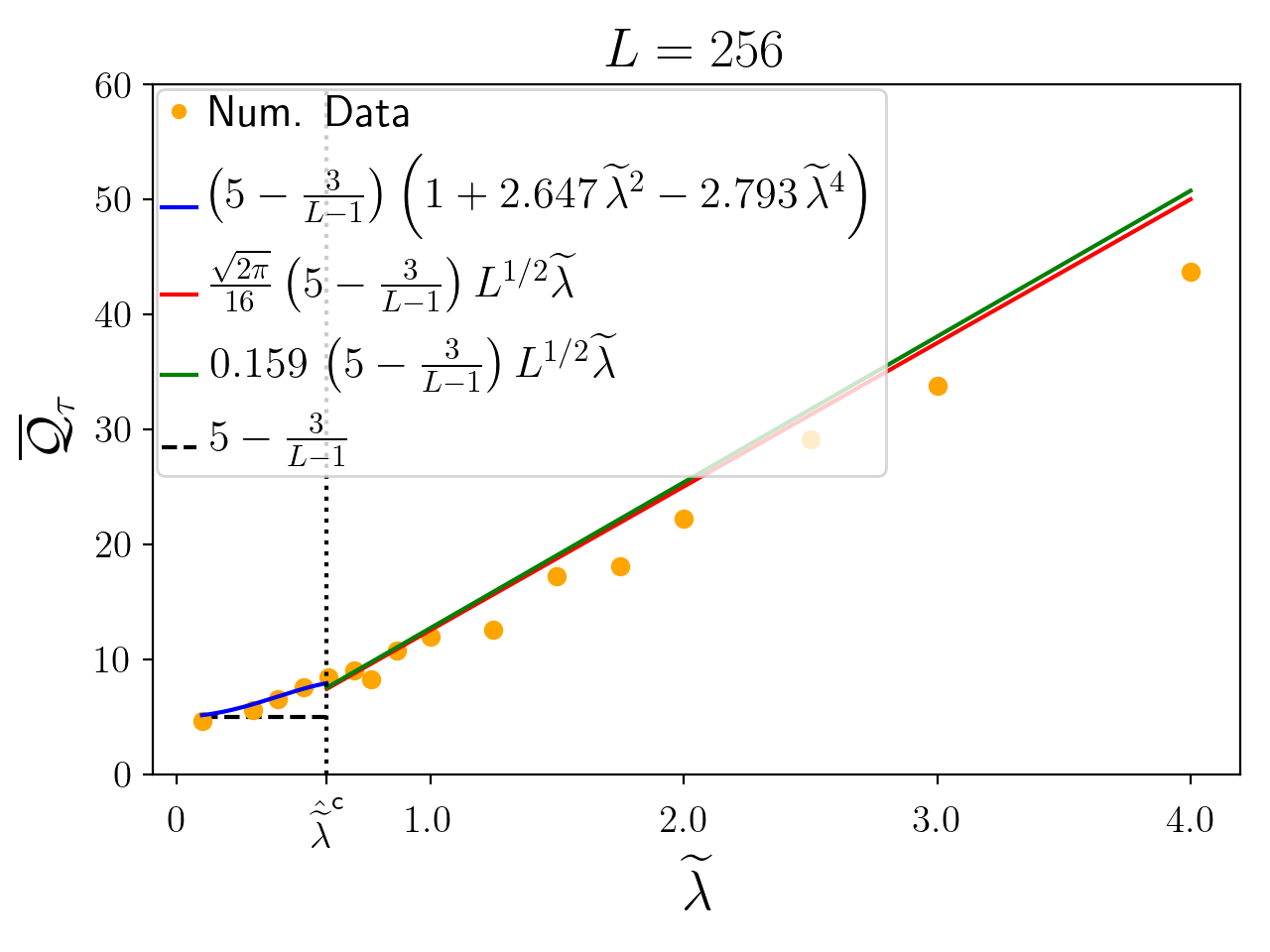}
\caption{}
\label{subfig:TUROfLambda_corr}
\end{subfigure}
\caption{Display of the numerical steady state values of the TUR product $\overline{\mathcal{Q}}_\tau$, obtained as time averages for times $t\geq\tau=10^4$ in dependence of $\wt{\lambda}$. The solid lines represent the theoretical predictions in the two scaling regimes. In (\subref{subfig:TUROfLambda}), these are taken from \eqref{eq:ResultMatchingConstant1} and \eqref{eq:DefNumFullTUR}, whereas in (\subref{subfig:TUROfLambda_corr}) we used \eqref{eq:NumTcKPZResult} and \eqref{eq:NumC0AmplitudeResult} to reevaluate \eqref{eq:ResultMatchingConstant1}.}
\label{fig:TUROfLambda}
\end{figure}
In Fig. \ref{fig:TUROfLambda} we show the TUR product in dependence of $\wt{\lambda}$. The values for $\overline{\mathcal{Q}}_\tau$ are obtained by calculating the temporal average in the KPZ stationary state of $\mathcal{Q}$ for times $t\geq\tau=10^4$. We expect the numerical data to follow the prediction in \eqref{eq:DefNumFullTUR}, which it does with good agreement as can be seen in Fig. \ref{fig:TUROfLambda}. Here the solid line below $\wt{\lambda}^\text{c}$ represents the perturbative result and is compared to the zero-order result depicted as the horizontal dashed line. In Fig. \ref{fig:TUROfLambda}(\subref{subfig:TUROfLambda}) we show for $\wt{\lambda}>\wt{\lambda}^\text{c}$, with $\wt{\lambda}^\text{c}$ from \eqref{eq:DefNumCritLambdaL256}, the theoretical predictions according to \eqref{eq:ResultMatchingConstant1} and \eqref{eq:DefNumFullTUR}, i.e., for the KPZ correlation time from \eqref{eq:KPZCorrelationTime}. On the other hand, Fig. \ref{fig:TUROfLambda}(\subref{subfig:TUROfLambda_corr}) displays the same theoretical predictions, reevaluated with the numerically obtained KPZ correlation time from \eqref{eq:NumTcKPZResult} and the ensuing $c_0$ from \eqref{eq:NumC0AmplitudeResult}. Using the KPZ correlation time $\wt{t}^\text{KPZ}_\text{c}$ from \eqref{eq:NumTcKPZResult} demands also a reevaluation of the critical value of the coupling parameter. Repeating the calculation of \eqref{eq:DefCriticalLeff} in \autoref{sec:ScalingVar}, we obtain for the numerically determined KPZ correlation time $\wt{t}^\text{KPZ}_\text{c}$ an effective critical coupling parameter of 
\begin{equation}
\hat{\lambda}_\text{eff}^\text{c}\approx9.47,\label{eq:DefCriticalLambda_corr}
\end{equation}
and thus for a system size of $L=256$
\begin{equation}
\hat{\wt{\lambda}}^\text{c}\approx0.592,\label{eq:DefNumCritLambda_corr}
\end{equation}
which is shown in Fig. \ref{fig:TUROfLambda}(\subref{subfig:TUROfLambda_corr}). As can be seen and is to be expected, using \eqref{eq:NumTcKPZResult}, \eqref{eq:NumC0AmplitudeResult} and \eqref{eq:DefNumCritLambda_corr} causes the two solid lines in Fig. \ref{fig:TUROfLambda}(\subref{subfig:TUROfLambda_corr}) for $\wt{\lambda}>\hat{\wt{\lambda}}^\text{c}$ to almost coincide and shrinks the jumps at the critical $\wt{\lambda}$ significantly in comparison to Fig. \ref{fig:TUROfLambda}(\subref{subfig:TUROfLambda}) (see also Fig. \ref{fig:TUROfT}(\subref{subfig:TUR_lambda4}), (\subref{subfig:TUR_lambda4_corr})). Another effect of introducing $\hat{\wt{\lambda}}^\text{c}$ is that for $\wt{\lambda}<\hat{\wt{\lambda}}^\text{c}$ the $4$th-order perturbation expansion from \eqref{eq:DefNumFullTUR} is cut off before it reaches its local maximum (as opposed to Fig. \ref{fig:TUROfLambda}(\subref{subfig:TUROfLambda})), which seems to be a physically more reasonable behavior.

\section{Conclusion}\label{sec:Conclusion}

We have given an analytical description of the thermodynamic uncertainty relation depending on the coupling strength of the KPZ non-linearity, see \eqref{eq:FullTUR}. In particular we showed that equal-time correlation functions, in the present case the steady state current $J$ and the entropy production rate $\sigma$, can be obtained exactly via functional integration using the known steady state probability density functional $p_\text{s}[h]$ of the $(1+1)$ dimensional KPZ equation, see \eqref{eq:ExactSSCurrent} and \eqref{eq:ExactSSSigma}, respectively. In case of $\text{var}[\Psi(t)]$ we extended the result from \cite{NiggemannSeifert2020} by calculating the next order of the perturbation expansion, valid in the EW scaling regime of the KPZ equation, see \eqref{eq:VarPerturbationResultDim}. Further, we approximated $\text{var}[\Psi(t)]$ in the KPZ scaling regime via a DRG approach and obtained an analytic expression in the transient KPZ scaling regime, which not only recovers the correct scaling form but moreover yields an explicit amplitude factor, see \eqref{eq:ResultVarPsiOfULongTimeLimit}. To our knowledge, this has not been done before. The knowledge of the general scaling behavior of $\text{var}[\Psi(t)]$ in a finite KPZ system, see Fig. \ref{fig:ScalingVarianceSchematic}, enables us to match the result from the DRG calculation in the transient KPZ regime to the stationary KPZ regime, see \eqref{eq:ResultMatchingConstant1}. We found that \eqref{eq:ResultMatchingConstant1} is in accordance with a result obtained via scaling arguments in \cite{KrugReview1997}, differing only in a numerical prefactor, the universal scaling amplitude $c_0$ in \cite{KrugReview1997}. The numerical value of this prefactor depends on the KPZ correlation time $\tcorr$ from \cite{KrugReview1997}, see \eqref{eq:KPZCorrelationTime}. During our numerical analysis, we found that for our data shown in \autoref{sec:CompNumSim} this theoretically predicted KPZ correlation time is roughly a factor of $2$ too small, see Tab. \ref{tab:TcOfLambda} and \eqref{eq:DefNumTcKPZ}. With the numerically obtained correlation time we reevaluated the calculation leading to the universal scaling amplitude $c_0$ and found that within the errorbars the result matches the theoretically predicted exact result in \cite{KrugReview1997}, see \eqref{eq:DefC0ScalingAmplitude}. We would like to emphasize that here the universal scaling amplitude $c_0$ has been determined, if only approximately, without any recourse to a particular model problem within the KPZ universality class and then relying on the universality hypothesis. Furthermore, we found good agreement between the numerical data and the theoretical predictions of the individual KPZ-TUR ingredients, namely $J$, $\sigma$ and $\text{var}[\Psi(t)]$, for arbitrary values of the coupling strength (see Figs. \ref{fig:Psi2Stot} - \ref{fig:VarKPZRegime}), as well as for the TUR product $\mathcal{Q}$ itself, both as a function of time (see Fig. \ref{fig:TUROfT}) and as a function of the coupling parameter (see Fig. \ref{fig:TUROfLambda}). In particular, we were able to describe the $\mathcal{Q}(\leff)$-behavior in the EW-scaling regime ($\leff\lessapprox\leff^\text{c}$) via a perturbation expansion up to $O(\leff^6)$ in the effective coupling parameter. It shows that in the weak coupling limit ($\leff\downarrow0$) $\mathcal{Q}(\leff)$ tends to $5$ from above. For the KPZ scaling regime we found asymptotically for $\leff^\text{c}\ll\leff$ a linear dependence of $\mathcal{Q}(\leff)$ on the effective coupling parameter. The perturbative description in the EW-regime is expected to hold for $\leff\uparrow\leff^\text{c}$, which is supported by the numerical results. However, it is not clear whether the DRG result, i.e., $\mathcal{Q}(\leff)\sim\leff$, remains valid for $\leff\downarrow\leff^\text{c}$, i.e., whether there are corrections to this linear behavior. While the numerical data do not indicate such corrections, their absence would imply that $\mathcal{Q}(\leff)$ is not smooth at $\leff=\leff^\text{c}$. Resolving this issue has to be left for future work.


\end{document}